\documentstyle[aps,prb,manuscript]{revtex}
\begin{document}
\newcommand{\bfsigma}{\mbox{{\boldmath $\sigma$}}}
\newcommand{\bfscrsigma}{\mbox{{\boldmath \scriptsize$\sigma$}}}
\draft
\def\dfrac#1#2{{\displaystyle{#1\over#2}}}
\title{Exact solutions for the periodic Anderson model in 2D: A 
Non-Fermi liquid state in normal phase.}
\author{Peter~Gurin and Zsolt~Gul\'acsi}
\address{
Department of Theoretical Physics, University of Debrecen, H-4010
Debrecen, Hungary }
\date{January, 2001}
\maketitle
\begin{abstract}
Presenting exact solutions for the two dimensional periodic Anderson model
with finite and nonzero on-site interaction $U > 0$, we are describing a
rigorous non-Fermi liquid phase in normal phase and 2D. This new state
emerges in multi-band interacting Fermi systems above half
filling, being generated by a flat band effect. The momentum distribution 
function $n_{\vec k}$ together with its derivatives of any order is 
continuous. The state possesses a well defined Fermi energy ($e_F$), but the 
Fermi momentum concept is not definable, so the Fermi surface in 
${\vec k}$-space is missing. The state emerges in the vicinity of a Mott 
insulating phase when lattice distortions are present, is highly degenerated
and paramagnetic. A gap is present at high $U$ in the density of low lying
excitations. During low lying excitations, quasi-particles are not created 
above the Fermi level, only the number of particles at $e_F$ increases. 
\end{abstract}

\section{Introduction}

The topic of non-Fermi liquid (NFL) behavior in $D \: > \: 1$ dimensions
and normal (non-symmetry broken) phase (NP) is currently of great interest
\cite{pbev1}. This is mainly due to the large amount of experimental
results, obtained in principle in the last decade, showing NFL behavior
in the NP of a variety of materials, including $D \: > \: 1$ dimensional
systems of great interest. Examples are: high $T_c$ superconductors
\cite{pbev2}, heavy-fermions \cite{pbev3}, layered systems \cite{pbev4},
quasi-one dimensional conductors, doped semiconductors, systems with
impurities, materials presenting proximity to metal-insulator transitions
\cite{pbev5}, etc. These results changed considerably our understanding of
interacting Fermi systems. Indeed, until recently, Fermi-liquid (FL) theory
seemed universally applicable to all sufficiently pure interacting Fermi
systems, and its main features even to dirty systems, provided that their
NP is not destroyed by a symmetry breaking process \cite{pbev1}. This
``dogma'' has been based on high precision experimental verifications in 
liquid $He 3$ and simple metals \cite{pbev6}. The concept of FL itself has 
been introduced by Landau many decades ago \cite{pbev7} (for a thorough 
discussion see \cite{pbev8}), and in principle has the meaning that in spite 
of the interactions, the low energy behavior can be well described within a 
picture of almost noninteracting quasi-particles. Formulated in rigorous terms 
\cite{pbev5,pbev9}, in a normal FL we have a one-to-one correspondence 
between the non-interacting
and interacting single-particle states (determined e.g., by a perturbation
theory convergent up to infinite orders). Furthermore, a quasi-particle pole is
present in the single-particle propagator that gives rise to a step-like
discontinuity of the momentum distribution function $n_{\vec k}$ at
the Fermi surface, whose position is specified by a sharp Fermi momentum value
${\vec k}_F$. The observation of NFL behavior in the materials presented
above polarized a huge intellectual effort in the last decade \cite{pbev10}
for the understanding of this new fermionic state. In this field the
theoretical interpretations are often based on multi-band models \cite{pmb1},
the presence of a some kind of gap in the NP being clearly
established in many cases and subject of intensive experimental \cite{pmb2}
and theoretical \cite{pmb1,pmb3} studies. However, despite the great
number of papers published in the field (see for example the references
cited in \cite{pbev1} or \cite{pbev9}), and the fact that the observed most
interesting and important normal NFL properties emerge in two
spatial dimensions (2D), (for example the normal phase of the
high $T_c$ superconductors), on the
theoretical side, for pure systems, the existence of a NFL state
in a NP has been proved exactly only in one dimensions (i.e. Luttinger
liquid \cite{pbev11}). The extension possibility of NFL-NP properties
to 2D has not been demonstrated rigorously up today. In fact, a rigorous  
theory of a NFL normal state in higher than one dimensions is missing.

Driven by these state of facts, we started to focus our attention on
possible NFL states using exact methods which are applicable in higher
than one dimensions as well \cite{mi1}. Based on the obtained results and a 
conjecture made by us\cite{mi2}, in this paper we are reporting the first 
rigorously derived NFL state in 2D. We deduced for this reason exact 
solutions for a real space version of the periodic Anderson model (PAM) in 2D.
The model is analyzed on a two dimensional square lattice, in case of
non-vanishing and finite Hubbard on-site repulsion $U$. In the paper in fact
two qualitatively different solutions are described: a completely 
localized and a non-localized one, which represents the
first exact results reported for PAM in 2D and finite $U$. 
The solutions are valid on two surfaces of the parameter space, i.e.
on restricted, but continuous and infinite regions of the $T=0$ phase
diagram, extended from the low $U$ to the high $U$ regions up to $U=\infty$
at $U>0$. 

The derived non-Fermi liquid state is given by a flat band effect 
in multi-band systems with more than half filling. The obtained properties are
extremely peculiar: the system in case of the described solution possesses a 
well defined Fermi energy $e_F$ in conditions in which the ${\vec k}_F$ 
Fermi momentum cannot be defined, and the $n_{\vec k}$ momentum distribution
function is continuous together with its derivatives of any order.
The system has also a natural built in gap, which is clearly present
in the density of low lying excitations at high $U$. The state is paramagnetic
and non-insulating. The gap symmetry is a possible symmetry allowed by the
described 2D lattice, and depends on the starting parameters of the system. 
The state emerges in the proximity of a Mott insulating phase when lattice 
distortions in the unit cell are present. During low lying excitations
quasi-particles are not created above the Fermi level, only the number of 
particles increases at $e_F$.

Concerning the flat band features (FBF), we mention
that such characteristics have been clearly observed
in different systems where strong electron interactions and strong correlation
effects play a main role. On numerical side, FBF are present for example in
results connected to 2D Hubbard model \cite{fb1,fb2,fb3},
or 2D $t-J$ model \cite{fb4}. Experimentally FBF are seen in angle-resolved
photo-emission (ARPES) data of high $T_c$ cuprates \cite{fb5,fb6}. For
layered systems ARPES often shows main bands without any sharp
characteristics in $n_{\vec k}$ \cite{fb7}, or give results interpreted via
FBF assumptions\cite{fb8}. Band structure calculations for these
systems often reflects a Fermi level positioned exactly at the bottom of a
conduction band with large effective mass around its minimum, below which a
gap is present \cite{pbev4}. We further wish to mention that connections
between superconductivity and FBF were also clearly pointed
out by Imada et al. \cite{fb9}, and FBF can be seen as well in experiments
related to heavy-fermion materials\cite{fb9a}. On the technological side, 
for example Lammert et al. \cite{fb10} have shown that squashing carbon 
nanotubes, FBF can be achieved around $e_F$, where a mismatch of nearly 
isoenergetic ${\vec k}$ states may have unexpected application possibilities.

The remaining part of the paper is constructed as follows:
Section II. presents in detail the analyzed model and the general form of the
deduced ground-state wave-functions. Section III. characterizes the obtained
solutions from the point of view of the ${\vec k}$ - space representation of 
the Hamiltonian and wave-vectors described, and Section IV. analyzes magnetic
properties of the system in the studied ground-states. Section V. describes a
completely localized insulating solution, Section VI. presents the new 
non-Fermi liquid state in normal phase and 2D, Section VII. summarizes 
the paper, and the Appendix, containing mathematical details, closes the 
presentation.

\section{The model and ground states deduced}

We are describing in this Section the model we use and the ground-states 
detected for it in restricted domains of the phase diagram.

\subsection{The model}

We are considering in this paper a 2D square lattice described by a two-band 
model whose Hamiltonian for the start is given in direct space, containing 
on-site repulsive interaction in one band. The starting point 
will be sufficiently general in order to give us the possibility to
characterize in detail the state we are presenting. However, the model 
contains also restrictions. Based on physical considerations, and denoting by
${\vec d}_n$ the positions of the $n$th neighbors of a given but arbitrary
lattice site, we are taking into consideration in this paper only the 
$n \: \leq \: 2$ (i.e. nearest, and next-nearest neighbor) contributions in 
the Hamiltonian. With these considerations, our starting Hamiltonian can be 
given as
\begin{eqnarray}
\hat H \: = \: \hat H_0 \: + \: \hat U \: , \quad
\hat H_0 \: = \: \hat T_c \: + \: \hat T_f \: + \: \hat V_0 \: + \: \hat V 
\: + \: \hat E_f \: ,
\label{ez01}
\end{eqnarray}
where the non-interacting terms have been denoted together by $\hat H_0$. 
With ${\bf i}$ denoting an arbitrary 2D lattice site position ${\vec r}_i$, 
the contributing terms in Eq.(\ref{ez01}) can be explicitly written as follows
\begin{eqnarray}
&& \hat T_c \: = \:
\sum_{{\bf j}, \sigma} \: \left( \: t_{c,x} \: \hat c^{\dagger}_{{\bf j},
\sigma} \: \hat c_{{\bf j} + {\bf x}, \sigma} \: + \: h.c. \: \right) \: + \: 
\sum_{{\bf j}, \sigma} \: \left( \: t_{c,y} \: \hat c^{\dagger}_{{\bf j}, 
\sigma} \: \hat c_{{\bf j} + {\bf y}, \sigma} \: + \: h.c. \: \right) \: +
\nonumber\\
&& \sum_{{\bf j}, \sigma} \: \left( \: t_{c,x+y} \: \hat c^{\dagger}_{{\bf j},
\sigma} \: \hat c_{{\bf j} + ( {\bf x} + {\bf y}), \sigma} \: + \: h.c. \:
\right) \: + \: \sum_{{\bf j}, \sigma} \: \left( \: t_{c,y-x} \: 
\hat c^{\dagger}_{{\bf j} + {\bf x}, \sigma} \: \hat c_{{\bf j} + {\bf y}, 
\sigma} \: + \: h.c. \: \right) \: ,
\label{ez04}
\end{eqnarray}
\begin{eqnarray}
&& \hat T_f \: = \:
\sum_{{\bf j}, \sigma} \: \left( \: t_{f,x} \: \hat f^{\dagger}_{{\bf j},
\sigma} \: \hat f_{{\bf j} + {\bf x}, \sigma} \: + \: h.c. \: \right) \: + \: 
\sum_{{\bf j}, \sigma} \: \left( \: t_{f,y} \: \hat f^{\dagger}_{{\bf j}, 
\sigma} \: \hat f_{{\bf j} + {\bf y}, \sigma} \: + \: h.c. \: \right) \: +
\nonumber\\
&& \sum_{{\bf j}, \sigma} \: \left( \: t_{f,x+y} \: \hat f^{\dagger}_{{\bf j},
\sigma} \: \hat f_{{\bf j} + ( {\bf y} + {\bf x} ), \sigma} \: + \: h.c. \:
\right) \: + \: \sum_{{\bf j}, \sigma} \: \left( \: t_{f,y-x} \: 
\hat f^{\dagger}_{{\bf j} + {\bf x}, \sigma} \: \hat f_{{\bf j} + {\bf y}, 
\sigma} \: + \: h.c. \: \right) \: ,
\label{ez05}
\end{eqnarray}
\begin{eqnarray}
&& \hat V_1 \: = \: \sum_{{\bf j}, \sigma} \: \left( \: V_{1,x}^{cf} \: 
\hat c^{\dagger}_{{\bf j}, \sigma} \: \hat f_{{\bf j} + {\bf x}, \sigma} \:
+ \: h.c. \: \right) \: + \: \sum_{{\bf j}, \sigma} \: \left( \: V_{1,x}^{fc} 
\: \hat f^{\dagger}_{{\bf j}, \sigma} \: \hat c_{{\bf j} + {\bf x}, \sigma} \:
+ \: h.c. \: \right) \: +  
\nonumber\\
&& \sum_{{\bf j}, \sigma} \: \left( \: V_{1,y}^{cf} \: 
\hat c^{\dagger}_{{\bf j}, \sigma} \: \hat f_{{\bf j} + {\bf y}, \sigma} \: +
\: h.c. \: \right) \: + \: \sum_{{\bf j}, \sigma} \: \left( \: V_{1,y}^{fc} 
\: \hat f^{\dagger}_{{\bf j}, \sigma} \: \hat c_{{\bf j} + {\bf y}, \sigma} \:
+ \: h.c. \: \right) \: ,
\label{ez06}
\end{eqnarray}
\begin{eqnarray}
&& \hat V_2 \: = \: \sum_{{\bf j}, \sigma} \: \left( \: V_{2,x+y}^{cf} \: 
\hat c^{\dagger}_{{\bf j}, \sigma} \: \hat f_{{\bf j} + ( {\bf x} + {\bf y} ),
\sigma} \: + \: h.c. \: \right) \: + \: \sum_{{\bf j}, \sigma} \: \left( \:
V_{2,x+y}^{fc} \: \hat f^{\dagger}_{{\bf j}, \sigma} \: \hat c_{{\bf j} + 
( {\bf x} + {\bf y} ), \sigma} \: + \: h.c. \: \right) \: +
\nonumber\\
&& \sum_{{\bf j}, \sigma} \: \left( \: V_{2,y-x}^{cf} \:
\hat c^{\dagger}_{{\bf j} + {\bf x}, \sigma} \: \hat f_{{\bf j} + {\bf y},
\sigma} \: + \: h.c. \: \right) \: + \: \sum_{{\bf j}, \sigma} \: \left( \:
V_{2,y-x}^{fc} \: \hat f^{\dagger}_{{\bf j} + {\bf x}, \sigma} \:
\hat c_{{\bf j} + {\bf y}, \sigma} \: + \: h.c. \: \right) \: ,
\label{ez07}
\end{eqnarray}
\begin{eqnarray}
\hat V_0 \: = \: \sum_{{\bf j}, \sigma} \: \left( \: V_0 \: 
\hat c^{\dagger}_{{\bf j}, \sigma} \: f_{{\bf j}, \sigma} \: + \:  h.c. \:
\right) \: , \quad
\hat E_f \: = \: E_f \: \sum_{{\bf j}, \sigma} \: \hat f^{\dagger}_{{\bf j},
\sigma} \: f_{{\bf j}, \sigma} \: , \quad
\hat U \: = \: U \: \sum_{\bf i} \: \hat n^{f}_{{\bf i}, \uparrow} \: 
\hat n^{f}_{{\bf i}, \downarrow} \: .
\label{ez08}
\end{eqnarray}
In these expressions ${\bf x}$ (${\bf y}$) represent elementary displacements
along a single bond of length equal with the lattice constant $a$ in $\vec{x}$
($\vec{y}$) direction, respectively. Taking into account different couplings 
along different lattice directions, we allow in fact the study of the system 
with distorted unit cell as well. The interaction term $\hat U$ is the usual 
Hubbard interaction, where 
$\hat n_{{\bf i}, \sigma}^f \: = \: {\hat f}_{{\bf i}, \sigma}^{\dagger} \:
{\hat f}_{{\bf i}, \sigma}$, is the particle number operator for 
$\hat f_{{\bf i}, \sigma}$ electrons.

Our starting ${\hat H}$ from Eq.(\ref{ez01}) represents a prototype of
an interacting two band system in 2D. It describes a square lattice containing
fermions distributed in two bands $b \: = \: c, \: f$. The kinetic energy 
term is in fact
$\hat{T} \: = \: \sum_{n=1}^2 \: \sum_{\langle n; {\bf i}, {\bf j} \rangle, b,
\sigma} \: ( \: t_{b, {\vec d}_n} \: {\hat b}^{\dagger}_{{\bf i}, \sigma} 
\: {\hat b}_{{\bf j}, \sigma} \: + \: h.c. ) \: ,$
where $t_{b, {\vec d}_n}$ are hopping amplitudes, $\sigma$ is the spin index, 
and ${\langle n; \: {\bf i} \: , \: {\bf j} \: \rangle}$ has the meaning of a
sum over bonds connecting $n$th neighbors, every bond being taken into
account once. The hybridization between the bands is composed from on-site
$\hat{V}_0$ and $n$th neighboring sites hybridizations $\hat{V} \: = \: \sum_{
n = 1, 2} \: \hat{V}_n $. As can be seen, in Eq.(\ref{ez01}) the non-local 
hybridization $\hat V \: = \: \hat V_1 \: + \: \hat V_2$ contains the 
nearest-neighbor ($\hat V_1$) and the next-nearest-neighbor ($\hat V_2$) 
contributions only, the $\hat V_n$, $n \geq 3$ long-range terms being 
considered negligibly small. The on-site energy $E_f$ for the second band 
fixes the relative position of the two bands, and finally, $U$ represents in 
fact the on-site Coulomb repulsion in the second band $b = f$, making the 
model interacting ( $U \: > \: 0$ ). We mention that $U$ is present on all
lattice sites. In order to help the reader in a better understanding of the
notations, we are presenting in Fig.1. the hopping amplitudes for 
$c$-electrons (A), and hybridization matrix elements for 
$\hat c^{\dagger}_i \hat f_j$ type transfers (B). The hopping amplitudes for
$f$-electrons, and the hybridization matrix elements for $\hat f^{\dagger}_i
\hat c_j$ type transfers are similar. For them, only the index $c$ has to be 
changed in $f$ in Fig.1.A., and the superscript $cf$ has to be replaced with
$fc$ in Fig.1.B. As given above, our Hamiltonian $\hat H$
can be considered as a 2D - PAM given in a direct space version, or a 2D 
two-band Hubbard model containing the contribution of the Hubbard $U$ only in
one band, the other band being non-interacting. 

We concentrate in this paper on a specific region of the phase diagram of the
model ${\cal P}_{\hat H}$. This region ${\cal P}_{\hat H}$ can be defined as
the domain of the parameter space where $\hat H$ can be expressed through 
elementary plaquette contributions in the form
$\hat{A}^{\dagger}_{I, \sigma} \: \hat{A}_{I, \sigma}$. Here, the plaquette
operator $\hat{A}_{I, \sigma}$ is build up from a linear combination of the
starting $\hat b$ annihilation fermionic operators acting on the elementary 
plaquette $I$. We are denoting the four positions available on a plaquette by 
$\: {\bf j} \: = \: 1,2,3,4 \: $, starting from the down-left corner and 
counting anti-clockwise (see the notation of sites from Fig.1.). 
The $I$th plaquette operator for a fixed spin $\sigma$ can be generally 
expressed as
$ \hat{A}_{I, \sigma} \: = \: \sum_{{\bf j}, b} \: \: a_{{\bf j}(I), b} \:
\hat{b}_{{\bf j}(I), \sigma}  \: ,$ where, because of the translation 
invariance, $a_{{\bf j}(I), b} \: = \: a_{{\bf j}, b}$ plaquette independence
will be considered for the emerging eight coefficients present in 
$\hat{A}_{I, \sigma}$, that have to be deduced. Taking the translational 
invariance explicitly into account, the plaquette operator acting on four 
corners of an elementary square plaquette $I$ becomes
\begin{eqnarray}
\hat A_{I,\sigma} \: & = & \: a_{1,f} \: {\hat f}_1 \: + \: a_{2,f} \:
{\hat f}_2 \: + \: a_{3,f} \: {\hat f}_3 \: + \: a_{4,f} \: {\hat f}_4 
\nonumber\\
& + & \: a_{1,c} \: {\hat c}_1 \: + \: a_{2,c} \: {\hat c}_2 \: + \:
a_{3,c} \: {\hat c}_3 \: + \: a_{4,c} \: {\hat c}_4 \: .
\label{ez1}
\end{eqnarray}
Starting from Eq.(\ref{ez1}), and taking into consideration periodic boundary 
conditions, the product $\hat A^{\dagger}_{I, \sigma} \: \hat A_{I, \sigma}$ 
summed over plaquettes for a fixed spin index $\sigma$ can be written as
\begin{eqnarray}
&& \sum_{I} \: \hat A^{\dagger}_{I, \sigma} \: \hat A_{I, \sigma} \: = \:
\left( \: \sum_{i = 1}^4 \: |a_{i, c}|^2 \: \right) \: \left( \: 
\sum_{{\bf j} = 1}^{N_{\Lambda}} \: {\hat c}^{\dagger}_{{\bf j}, \sigma} \:
{\hat c}_{{\bf j}, \sigma} \: \right) \: + \: \left( \: \sum_{i = 1}^4 \:
|a_{i,f}|^2 \: \right) \: \left( \: \sum_{{\bf j} = 1}^{N_{\Lambda}} \: 
{\hat f}^{\dagger}_{{\bf j}, \sigma} \: {\hat f}_{{\bf j}, \sigma} \:
\right) \: + 
\nonumber\\
&& \sum_{{\bf j} = 1}^{N_{\Lambda}} \: \left\{ \: \left[ \: \left( \:
a_{1,c}^{*} \: a_{2,c} \: + \: a_{4,c}^{*} \: a_{3,c} \: \right) \:
{\hat c}_{{\bf j}, \sigma}^{\dagger} \: {\hat c}_{{\bf j} + {\bf x}, \sigma} 
\: + \: h.c. \: \right] \: + \right. 
\nonumber\\
&& \left[ \: \left( \: a_{1,c}^{*} \: a_{4,c} \:
+ \: a_{2,c}^{*} \: a_{3,c} \: \right) \: {\hat c}_{{\bf j}, \sigma}^{\dagger}
\: {\hat c}_{{\bf j} + {\bf y}, \sigma} \: + \: h.c. \: \right] \: +
\nonumber\\
&& \left[ \: \left( \: a_{1,f}^{*} \: a_{2,f} \: + \: a_{4,f}^{*} \: a_{3,f} 
\: \right) \: {\hat f}_{{\bf j}, \sigma}^{\dagger} \: {\hat f}_{{\bf j} + 
{\bf x}, \sigma} \: + \: h.c. \: \right] \: + 
\nonumber\\
&& \left[ \: \left( \:
a_{1,f}^{*} \: a_{4,f} \: + \: a_{2,f}^{*} \: a_{3,f} \: \right) \:
{\hat f}_{{\bf j}, \sigma}^{\dagger} \: {\hat f}_{{\bf j} + {\bf y}, \sigma} 
\: + \: h.c. \: \right] \: +
\nonumber\\
&& \left[ \: a_{1,c}^{*} \: a_{3,c} \: {\hat c}^{\dagger}_{{\bf j}, \sigma} \:
{\hat c}_{{\bf j} + ( {\bf x} + {\bf y} ), \sigma} \: + \: h.c. \: \right] \:
+ \: \left[ \: a_{2,c}^{*} \: a_{4,c} \: {\hat c}^{\dagger}_{{\bf j}, \sigma} 
\: {\hat c}_{{\bf j} + ( {\bf y} - {\bf x} ), \sigma} \:  
+ \: h.c. \: \right] \: +
\nonumber\\
&& \left[ \: a_{1,f}^{*} \: a_{3,f} \: {\hat f}^{\dagger}_{{\bf j}, \sigma} 
\: {\hat f}_{{\bf j} + ( {\bf x} + {\bf y} ), \sigma} \:
+ \: h.c. \: \right] \: + \: \left[ \: a_{2,f}^{*} \: a_{4,f} \:
{\hat f}^{\dagger}_{{\bf j}, \sigma} \: {\hat f}_{{\bf j} + ( {\bf y} - 
{\bf x} ), \sigma} \: + \: h.c. \: \right] \: +
\nonumber\\
&& \left[ \: \left( \: a^{*}_{1,c} \: a_{2,f} \: + \: a^{*}_{4,c} \: a_{3,f} 
\: \right) \: {\hat c}^{\dagger}_{{\bf j}, \sigma} \: {\hat f}_{{\bf j} + 
{\bf x}, \sigma} \: + \: h.c. \: \right] \: + 
\nonumber\\
&& \left[ \: \left( \: 
a^{*}_{1,f} \: a_{2,c} \: + \: a^{*}_{4,f} \: a_{3,c} \: \right) \: 
{\hat f}^{\dagger}_{{\bf j}, \sigma} \: {\hat c}_{{\bf j} + {\bf x}, \sigma} 
\: + \: h.c. \: \right] \: +
\nonumber\\
&& \left[ \: \left( \: a^{*}_{1,c} \: a_{4,f} \: + \: a^{*}_{2,c} \: a_{3,f}
\: \right) \: {\hat c}^{\dagger}_{{\bf j}, \sigma} \: {\hat f}_{{\bf j} + 
{\bf y}, \sigma} \: + \: h.c. \: \right] \: + 
\nonumber\\
&& \left[ \: \left( \: 
a^{*}_{1,f} \: a_{4,c} \: + \: a^{*}_{2,f} \: a_{3,c} \: \right) \: 
{\hat f}^{\dagger}_{{\bf j}, \sigma} \: {\hat c}_{{\bf j} + {\bf y}, \sigma} 
\: + \: h.c. \: \right] \: +
\nonumber\\
&& \left[ \: a_{1,c}^{*} \: a_{3,f} \: {\hat c}^{\dagger}_{{\bf j}, \sigma} 
\: {\hat f}_{{\bf j} + ( {\bf x} + {\bf y} ), \sigma} \: + \: h.c. \: \right]
\: + \: \left[ \: a_{1,f}^{*} \: a_{3,c} \: {\hat f}^{\dagger}_{{\bf j},
\sigma} \: {\hat c}_{{\bf j} + ( {\bf x} + {\bf y} ), \sigma} \: + \:
h.c. \: \right] \: +
\nonumber\\
&& \left[ \: a_{2,c}^{*} \: a_{4,f} \: {\hat c}^{\dagger}_{{\bf j}, \sigma} 
\: {\hat f}_{{\bf j} + ( {\bf y} - {\bf x} ), \sigma} \: + \: h.c. \: 
\right] \: + \: \left[ \: a_{2,f}^{*} \: a_{4,c} \: {\hat f}^{\dagger}_{
{\bf j}, \sigma} \: {\hat c}_{{\bf j} + ( {\bf y} - {\bf x} ), \sigma} \:
+ \: h.c. \: \right] \: +
\nonumber\\
&& \left.
\left[ \: \left( \: \sum_{i=1}^4 \: a_{i,c}^{*} \: a_{i,f} \: \right) \:
\left( \: \sum_{{\bf j} = 1}^{N_{\Lambda}} \: {\hat c}^{\dagger}_{{\bf j},
\sigma} \: {\hat f}_{{\bf j}, \sigma} \: \right) \: + \: h.c. \:
\right] \: \right\} \: . 
\label{ez2}
\end{eqnarray}
In Eq.(\ref{ez2}), $N_{\Lambda}$ denotes the number of lattice sites.
A comparison of Eq.(\ref{ez01}) and Eq.(\ref{ez2}) shows that $\hat H_0$
can be expressed via the $\hat A_{I, \sigma}$ operators as
$ \hat H_0 \: = \: - \: \sum_{I, \sigma} \: \hat A^{\dagger}_{I, \sigma} \:
A_{I, \sigma} \: + \: K \: \hat N $, if the following conditions are satisfied
\begin{eqnarray}
&& - \: t_{c,x} \: = \: a^{*}_{1,c} \: a_{2,c} \: + \: a^{*}_{4,c} \:
a_{3,c} \: , \quad
- \: t_{c,y} \: = \: a^{*}_{1,c} \: a_{4,c} \: + \: a^{*}_{2,c} \: 
a_{3,c} \: ,
\nonumber\\
&& - \: t_{f,x} \: = \: a^{*}_{1,f} \: a_{2,f} \: + \: a^{*}_{4,f} \:
a_{3,f} \: , \quad
- \: t_{f,y} \: = \: a^{*}_{1,f} \: a_{4,f} \: + \: a^{*}_{2,f} \:
a_{3,f} \: ,
\nonumber\\
&& - \: t_{c,x+y} \: = \: a^{*}_{1,c} \: a_{3,c} \: , \quad
- \: t_{c,y-x} \: = \: a^{*}_{2,c} \: a_{4,c} \: , \quad
- \: t_{f,x+y} \: = \: a^{*}_{1,f} \: a_{3,f} \: , \quad
- \: t_{f,y-x} \: = \: a^{*}_{2,f} \: a_{4,f} \: ,
\nonumber\\
&& - \: V_{1,x}^{cf} \: = \: a^{*}_{1,c} \: a_{2,f} \: + \: a^{*}_{4,c} \:
a_{3,f} \: , \quad 
- \: V_{1,x}^{fc} \: = \: a^{*}_{1,f} \: a_{2,c} \: + \: a^{*}_{4,f} \:
a_{3,c} \: , 
\nonumber\\
&& - \: V_{1,y}^{cf} \: = \: a^{*}_{1,c} \: a_{4,f} \: + \: a^{*}_{2,c} \:
a_{3,f} \: , \quad
- \: V_{1,y}^{fc} \: = \: a^{*}_{1,f} \: a_{4,c} \: + \: a^{*}_{2,f} \:
a_{3,c} \: , 
\nonumber\\
&& - \: V_{2,x+y}^{cf} \: = \: a^{*}_{1,c} \: a_{3,f} \: , \quad
- \: V_{2,x+y}^{fc} \: = \: a^{*}_{1,f} \: a_{3,c} \: , 
\nonumber\\
&& - \: V_{2,y-x}^{cf} \: = \: a^{*}_{2,c} \: a_{4,f} \: , \quad
- \: V_{2,y-x}^{fc} \: = \: a^{*}_{2,f} \: a_{4,c} \: ,
\nonumber\\
&& - \: V_0 \: = \: \sum_{i = 1}^4 \: a^{*}_{i,c} \: a_{i,f} \: , \quad
K \: = \: \sum_{i = 1}^{4} \: |a_{i,c}|^2 \: , 
\label{ez4}
\end{eqnarray}
\begin{eqnarray}
E_f \: = \: K \: - \: \sum_{i = 1}^4 \: |a_{i,f}|^2 \: .
\label{ez401}
\end{eqnarray}
In the expression of $\hat H_0$ the operator $\hat N \: = \: \sum_{{\bf i}, b,
\sigma} \: \hat n_{{\bf i}, \sigma}^b$ represents the total particle number 
operator. Based on Eq.(\ref{ez1}), it can be observed that
$ \hat A^{\dagger}_{I, \sigma} \: \hat A_{I, \sigma} \: + \:
\hat A_{I, \sigma} \: \hat A^{\dagger}_{I, \sigma} \: = \: \sum_{i = 1}^4 
\: [ \: |a_{i,c}|^2 \: + \: |a_{i,f}|^2 \: ] \: = \: 2 \: K \: - \: E_f \: ,$
so the Hamiltonian $\hat H_0$ can be written as
\begin{eqnarray}
\hat H_0 \: = \: \sum_{I, \sigma} \: \hat A_{I, \sigma} \: 
\hat A^{\dagger}_{I, \sigma} \: + \: K \: \hat N \: - \: 2 \: [ \: 2 \: K \:
- \: E_f\: ] \: N_{\Lambda} \: .
\label{ez6}
\end{eqnarray}

In the presence of the interaction, besides $\hat H_0$ from 
Eq.(\ref{ez6}), we have to take into consideration the Hubbard term
$\hat U$ as well. However, we may observe that $\hat U$ can be exactly 
transformed as
\begin{eqnarray}
\hat U \: = \: U \hat P' \: + \: U \: \sum_{{\bf i}, \sigma} \: 
\hat n^f_{{\bf i}, \sigma} \: - \: U \: N_{\Lambda} \: ,
\label{ez7}
\end{eqnarray}
where $\hat P' \: = \: \sum_{\bf i} \: \hat P'_{\bf i}$ and 
$ \hat P'_{\bf i} \: = \: ( \: 1 \: - \: {\hat n}_{{\bf i}, 2, \uparrow} \:
- \: {\hat n}_{{\bf i}, 2, \downarrow} \: + \: {\hat n}_{{\bf i}, 2, 
\uparrow} \: {\hat n}_{{\bf i}, 2, \downarrow} \: ) \: .$
In the decomposition presented in Eq.(\ref{ez7}) the ${\hat P'}$ operator
is a positive semidefinite operator. The reason for this is simple. 
${\hat P'}_{{\bf i}}$ applied to a wave function gives one if on the 
${\bf i}$ site there are no $f$ electrons present, and gives zero, if on 
the site ${\bf i}$ there is at least one $f$ electron present. As a 
consequence, ${\hat P'}$ representing a sum of non-negative numbers, it is a 
positive semidefinite operator.

We have further to observe that in Eq.(\ref{ez7}) the $ U \: \sum_{{\bf i},
\sigma} \: \hat n^f_{{\bf i}, \sigma}$ term simply renormalizes 
the $E_f$ contribution from the Hamiltonian. Keeping this information, 
introducing the notations $ \tilde E_f \: = \: E_f \: + \: U \: ,$  and 
$ \hat G \: = \: \sum_{I, \sigma} \: \hat A_{I, \sigma} \:
\hat A^{\dagger}_{I, \sigma} \: ,$ the starting Hamiltonian from 
Eq.(\ref{ez01}) can be written as
\begin{eqnarray}
\hat H \: = \: [ \: \hat G \: + \: U \: \hat P' \: ] \: - \: [ \: U \:
N_{\Lambda} \: + \: 2 \: ( \: 2 \: K \: - \: \tilde E_f \: ) \: N_{\Lambda} 
\: - \: K \: \hat N \: ] \: .
\label{ez11}
\end{eqnarray}
The decomposition presented in Eq.(\ref{ez11}) is valid if the conditions
presented in Eq.(\ref{ez4}) are satisfied, and in Eq.(\ref{ez401}) the $E_f$
value is replaced by $\tilde E_f$, where
\begin{eqnarray}
\tilde E_f \: = \: K \: - \: \sum_{i = 1}^4 \: |a_{i,f}|^2 \: .
\label{ez401a}
\end{eqnarray}
We mention that from mathematical point of view, the parameter space region 
${\cal P}_{\hat H}$ is given by the solutions of the system of 
equations Eqs.(\ref{ez4},\ref{ez401a}). If this system of equations admits
solutions for the coefficients $a_{i,b}$ from Eq.(\ref{ez1}), we are situated
inside ${\cal P}_{\hat H}$.

From Eq.(\ref{ez11}) can be seen that in conditions presented by
Eqs.(\ref{ez4},\ref{ez401a}), the analyzed Hamiltonian from Eq.(\ref{ez01}) 
can be written as
\begin{eqnarray}
\hat H \: = \: \hat P \: + \: E^U_0 \: ,
\label{ez12}
\end{eqnarray}
where the positive semidefinite operator
$\hat P$ and the constant number $E^U_0$ are given by
$\hat P \: = \: \hat G \: + \: U \: \hat P' \: ,$ and
\begin{eqnarray}
E^U_0 \: = \: K \: N \: + \: ( \: U \: + \: 2 \: E_f \: - \: 4 \: K \:
) \: N_{\Lambda} \: ,
\label{ez14}
\end{eqnarray}
where $N$ being the eigenvalue of $\hat N$, represents the number of particles
within the system. 

\subsection{The detected ground-states}

For the Hamiltonian from Eq.(\ref{ez12}), $\hat P$ 
being a positive semidefinite operator, the ground-state energy is 
$E_g^U = E^U_0$, and the ground-state wave function is that 
$| \: \Psi_g^U \: \rangle$, for which we have 
$\hat P \: | \: \Psi^U_g \: \rangle \: = \: 0$. The interesting aspect from 
the mathematical side of the problem is that $| \: \Psi^U_g \: \rangle$ can 
be explicitly expressed in the concentration range 
situated at and above $3/4$ filling, obtained in this way a many-body,
fully quantum mechanical solution in the interacting case and 2D. For this, 
we have to take into consideration that starting from the definition 
presented in Eq.(\ref{ez1}), for plaquette operators we have
$ \hat A^{\dagger}_{I, \sigma} \: \hat A^{\dagger}_{I', \sigma'} \: + \:
\hat A^{\dagger}_{I', \sigma'} \: \hat A^{\dagger}_{I, \sigma} \: = \: 0 \: $.
Furthermore, the number of elementary plaquettes $I$ from the system is equal
with the number of lattice sites, and the product $ \hat F^{(3)} \: = \:
\prod_{\bf i} \: \hat F^{\dagger}_{{\bf i}, f}$ based on the operator
(see also Eq.(\ref{ez16a}))
\begin{eqnarray}
\hat F_{{\bf i}, f} \: = \: \left( \: \alpha_{{\bf i}, \uparrow} \: 
\hat f_{{\bf i}, \uparrow} \: + \: \alpha_{{\bf i}, \downarrow} \:
\hat f_{{\bf i}, \downarrow} \: \right)
\label{ez16}
\end{eqnarray}
containing arbitrary $\alpha_{{\bf i}, \sigma}$ constants, creates an $f$ 
electron on every site of the lattice. As a consequences, for example at
$3/4$ filling (i.e. $N = 3 N_{\Lambda}$), the ground-state (not normalized)
wave function $| \: \Psi^U_g \: \rangle$ becomes the ordered product
\begin{eqnarray}
| \: \Psi^U_g \: \rangle \: = \: \prod_{\beta = 1}^{3} \: \hat F^{(\beta)} \:
| \: 0 \: \rangle \: ,
\label{ez17}
\end{eqnarray}
where
\begin{eqnarray}
\hat F^{(1)} \: = \: \prod_{I} \: \hat A^{\dagger}_{I, \uparrow} \: , \quad
\hat F^{(2)} \: = \: \prod_{I} \: \hat A^{\dagger}_{I, \downarrow} \: ,
\label{ez18}
\end{eqnarray}
and $| \: 0 \: \rangle$ being the bare vacuum with no fermions present. 
Indeed, because of the
$\hat A^{\dagger}_{I,\sigma} \: \hat A^{\dagger}_{I,\sigma} \: = \: 0$ 
property, we have $\hat G \: | \: \Psi^U_g \: \rangle \: = \: 0$ 
given by the $\hat F^{(1)} \: \hat F^{(2)}$ product in Eq.(\ref{ez17}), while
$\hat F^{(3)}$ introducing on every lattice site one electron, obliges
$| \: \Psi^U_g \: \rangle$ to have at least one $f$ particle on each site, 
preserving $\hat P' \: | \: \Psi^U_g \: \rangle \: = \: 0$ as explained 
below Eq.(\ref{ez7}). 

We would like to underline here that the $\alpha_{{\bf i}, \sigma}$ 
coefficients in Eq.(\ref{ez16}) are completely arbitrary, so the most general
form of the ground-state is in fact a linear combination of components present
in Eq.(\ref{ez17}) taken over all possible combinations of all possible 
$\alpha_{{\bf i}, \sigma}$ values. This means that in the most 
general case, the $\hat F_{{\bf i},f}$ operator (see Eq.(\ref{ez16}))
contained in $\hat F^{(3)}$, has the form  
\begin{eqnarray}
\hat F_{{\bf i}, f} \: = \: \sum_{ \{ \alpha_{{\bf i}, \sigma} \} } \:
g_{ \{ \alpha_{{\bf i}, \sigma} \} } \: \left( \: \alpha_{{\bf i}, \uparrow} 
\: \hat f_{{\bf i}, \uparrow} \: + \: \alpha_{{\bf i}, \downarrow} \: 
\hat f_{{\bf i}, \downarrow} \: \right) \: ,
\label{ez16a}
\end{eqnarray}
where $g_{ \{ \alpha_{{\bf i},\sigma} \} }$ represents numerical coefficients
(connected to fixed $\{ \alpha_{{\bf i}, \sigma} \}$ sets) restricted only 
by the normalization to unity of the whole wave function.

The solution for the ground-state can be also written for the system doped
above $3/4$ filling with arbitrary $ 1 \: \leq \: n_r \: < \: N_{\Lambda}$ 
number of electrons. For this reason we must define the operator
\begin{eqnarray}
\hat F^{(4)}_{\bf i} \: = \: \epsilon_{\uparrow} \: ( \: 
\hat f^{\dagger}_{{\bf i}, \uparrow} \: + \: e^{ i \phi_c } \: 
\hat c^{\dagger}_{{\bf i}, \uparrow} \: ) \: + \: \epsilon_{\downarrow} \: 
( \: \hat f^{\dagger}_{{\bf i}, \downarrow} \: + \: e^{ i \phi_c } \:
\hat c^{\dagger}_{{\bf i}, \downarrow} \: ) \: ,
\label{ez19}
\end{eqnarray}
which allows us to introduce randomly $n_r$ additional electrons in the
sistem above $3 \: N_{\Lambda}$ via
\begin{eqnarray}
\hat F^{(4)} \: = \: \sum_{ \{ {\bf i} \} } \: C_{ \{ {\bf i} \} } \:
\prod_{\bf i}^{n_r} \: \hat F^{(4)}_{\bf i} \: . 
\label{ez20}
\end{eqnarray}
In Eq.(\ref{ez20}), $\prod_{\bf i}^{n_r}$ represents an ordered product 
containing $n_r$ arbitrary chosen lattice sites taken as a possible 
combination of $n_r$ sites from $N_{\Lambda}$ possibilities, and 
$C_{ \{ {\bf i} \} }$ are numerical coefficients. As a consequence, in the 
doped case with $1 \: \leq \: n_r \: < \: N_{\Lambda}$, the ground-state 
wave function becomes
\begin{eqnarray}
| \: \Psi^{U}_{g,d} \: \rangle \: = \: \prod_{\beta = 1}^{4} \: 
\hat F^{\beta} \: | \: 0 \: \rangle \: .
\label{ez21}
\end{eqnarray}
We mention that along this paper we will restrict our study to the
$N \: \geq \: 3 \: N_{\Lambda}$ (i.e. $n_r \: \geq \: 0$) case.

It is extremely important to mention that the presented ground-state wave
functions in Eqs. (\ref{ez17}, \ref{ez21}) are valid only in interacting
($U \: > \: 0$) case, and cannot be perturbatively obtained from the 
$U \: = \: 0$ noninteracting limit. The reason for this is that at 
$U \: = \: 0$, as explained below Eq.(\ref{ez18}), the 
$\hat P \: | \Psi^{U = 0}_g \: \rangle \: = \: 0$ property is 
entirely given by the $\hat F^{(1)} \: \hat F^{(2)}$ product, the 
$\hat F^{(3)}$ operator being completely arbitrary. In the noninteracting case
the operator $\hat P$ reduces to $\hat G$, so we obtain inside 
${\cal{P}}_{\hat H}$ at $U \: = \: 0$ the equality 
$\hat G \: | \: \Psi^{0}_g \: \rangle \: = \: 0$ with the 
ground-state wave function 
\begin{eqnarray}
| \: \Psi^{0}_g \: \rangle \: = \: \left[ \: 
\prod_{\beta = 1}^{2} \: \hat F^{\beta} \: \right] \: \hat Q \: | \: 0 \:
\rangle \: , 
\label{Ueq0vf}
\end{eqnarray}
where $\hat Q$ is an arbitrary operator. As it can be seen, the 
concrete expression of the $\hat F^{(3)}$ operator is determined exactly by 
the nonzero $U \: > \: 0$ value of the interaction (see below 
Eq.(\ref{ez18})). Based on these characteristics mentioned,
we would like to underline that is no way to re-obtain (together 
with all expectation values that it gives) $| \: \Psi^{U}_g \: \rangle$ from 
$| \: \Psi^{0}_g \: \rangle$ (or vice versa) in the $U \: \to \: 0$ limit.

The remaining part of the paper is devoted to the study of the physical
properties of the $| \: \Psi^U_g \: \rangle$ wave function. We stress that 
depending on different solutions allowed by the system of equations 
Eqs.(\ref{ez4},\ref{ez401a}), the ground-state wave function given 
mathematically in Eq.(\ref{ez17}) describes from physical point of view 
even qualitatively different ground-states, which will be analyzed below.

\section{The $k$-space representations}
 
First of all, we would like to understand the physical background of the 
Hamiltonian form presented in Eq.(\ref{ez12}). We remember, 
that $\hat H$ from Eq.(\ref{ez12}) represents an exact
representation of the starting Hamiltonian from Eq.(\ref{ez01}) in conditions
in which Eqs.(\ref{ez4}, \ref{ez401a}) are valid (and admit solutions).

\subsection{The Fourier transform for $\hat H$}

In order to answer this question, let us transform the starting Hamiltonian
from Eq.(\ref{ez01}) in ${\vec k}$-space. In order to do this systematically,
let us first concentrate on $\hat H_0$. Denoting by ${\vec k}$ the 
two-dimensional reciprocal space vector, and using for the operators 
the $b \: = \: c, \: f$ notation together with the Fourier sum
$\hat b_{\bf j} \: = \: \sum_{\vec k} \: \hat b_{\vec k} \: exp \: \left[ \:
- \: i \: {\vec k} \cdot {\vec r}_{\bf j} \: \right] \: ,$
the kinetic energy terms becomes
$ \hat T_{b} \: = \: \sum_{{\vec k}, \sigma} \: \varepsilon_{{\vec k},
\sigma}^{b} \: \hat b^{\dagger}_{{\vec k}, \sigma} \: \hat b_{{\vec k},
\sigma} $, where
\begin{eqnarray}
&&\varepsilon_{{\vec k}, \sigma}^{b}  \: = \:  
\left( \: t_{b,x} \: e^{ - \: i \: {\vec k} \cdot {\vec x} } \: + \: 
t^{*}_{b,x} \: e^{ + \: i \: {\vec k} \cdot {\vec x} }  \: \right) \: + \:
\left( \: t_{b,y} \: e^{ - \: i \: {\vec k} \cdot {\vec y} } \: + \:
t^{*}_{b,y} \: e^{ + \: i \: {\vec k} \cdot {\vec y} } \: \right) \: + 
\nonumber\\
&&\left( \: t_{b,x+y} \: e^{ - \: i \: {\vec k} \cdot ( {\vec x} + {\vec y} )}
\: + \: t^{*}_{b,x+y} \: e^{ + \: i \: {\vec k} \cdot ( {\vec x} + {\vec y} )}
\: \right) \: + \: \left( \: t_{b,y-x} \: e^{ - \: i \: {\vec k} \cdot 
( {\vec y} - {\vec x} )} \: + \: t^{*}_{b,y-x} \: e^{ + \: i \: 
{\vec k} \cdot ( {\vec y} - {\vec x} ) } \: \right) \: .
\label{ez26}
\end{eqnarray}
Concerning the notations, we mention that using ${\vec d}_n$ introduced at the
beginning of Sec. II., defining $t_{b, {\vec d}_n} \: = \: | \: 
t_{b,{\vec d}_n} \: | \: exp ( \: i \: \phi_{t, b, {\vec d}_n} \: )$, we 
simply have $\varepsilon_{{\vec k}, \sigma}^{b} \: = \: 2 \: 
\sum_{{\vec d}_n} \: | \: t_{b, {\vec d}_n} \: | \: \cos ( \: 
\phi_{t, b, {\vec d}_n} \: - \: {\vec k} \cdot {\vec d}_n \: )$.

For the on-site energy at the $f$ level we simply obtain
$\hat E_f \: = \: E_f \: \sum_{{\vec k}, \sigma} \: \hat f^{\dagger}_{
{\vec k}, \sigma} \: \hat f_{{\vec k}, \sigma} \: .$
Introducing the notation $\hat T_f' \: = \: \hat T_f \: + \: \hat E_f$ 
together with the definitions $ \epsilon_{{\vec k}, \sigma}^c \: = \: 
\varepsilon_{{\vec k}, \sigma}^c, \quad \epsilon_{{\vec k}, \sigma}^f \: =
\: E_f \: + \: \varepsilon_{{\vec k}, \sigma}^f \: ,$ we simply have at
the level of Fourier transforms
$\hat T_c \: + \: \hat T_f' \: = \: \sum_{b,{\vec k}, \sigma} \:
\epsilon_{{\vec k}, \sigma}^b \: \hat b^{\dagger}_{{\vec k}, \sigma} \:
\hat b_{{\vec k}, \sigma} $ (note the difference between $\varepsilon^f_{
{\vec k}, \sigma}$ and $\epsilon^f_{{\vec k}, \sigma}$). 

In the case of the hybridization, the on-site term becomes
$ \hat V_0 \: = \: \sum_{{\vec k}, \sigma} \: ( \: V_0 \: 
\hat c^{\dagger}_{{\vec k}, \sigma} \: \hat f_{{\vec k}, \sigma} \: + 
\: h.c. \: ) \: .$
For the non-local hybridization $\hat V$, as shown in Sec.II.,
we have $\hat V \: = \: \hat V_1 \: + \: \hat V_2$, where, for 
$n \: = \: 1 , 2$ we obtain
$\hat V_n \: = \: \sum_{{\vec k}, \sigma} \: \left( \: V_{n, {\vec k}} \:
\hat c^{\dagger}_{{\vec k}, \sigma} \: \hat f_{{\vec k}, \sigma} \: + 
h.c. \: \right) \: .$ The hybridization matrix elements in these expressions 
for $n \: = \: 1, 2$, are given by
\begin{eqnarray}
&& V_{1, {\vec k}} \: = \: \left( \: V_{1,x}^{cf} \: e^{ - \: i \: 
{\vec k} \cdot {\vec x} } \: + \: V_{1,x}^{fc^*} \: e^{ + \: i \: 
{\vec k} \cdot {\vec x} } \: \right) \: + \: \left( \: V_{1,y}^{cf} \:
e^{ - \: i \: {\vec k} \cdot {\vec y} } \: + \: V_{1,y}^{fc^*} \:
e^{ + \: i \: {\vec k} \cdot {\vec y} } \: \right) \: ,
\nonumber\\
&& V_{2, {\vec k}} \: = \: \left( \: V_{2,x+y}^{cf} \: e^{ - \: i \: {\vec k} 
\cdot ( {\vec x} + {\vec y} ) } \: + \: V_{2,x-y}^{fc^*} \: e^{ + \: i \:
{\vec k} \cdot ( {\vec x} + {\vec y} ) } \: \right) \: + 
\nonumber\\
&& \left( \:
V_{2,y-x}^{cf} \: e^{ - \: i \: {\vec k} \cdot ( {\vec y} - {\vec x} ) } \:
+ \: V_{2,y-x}^{fc^*} \: e^{ + \: i \: {\vec k} \cdot ( {\vec y} - {\vec x} )
} \:  \right) \: .
\label{ez31}
\end{eqnarray}
Introducing the notation $V_{\vec k} \: = \: V_0 \: + \: V_{1, {\vec k}} \:
+ \: V_{2, {\vec k}}$, the total hybridization can be given as
$\hat V_0 \: + \: \hat V \: = \: \sum_{{\vec k}, \sigma} \: \left( \:
V_{\vec k} \: \hat c^{\dagger}_{{\vec k}, \sigma} \: \hat f_{{\vec k},
\sigma} \: + \: h.c. \: \right) \: ,$ and for $\hat H_0$ we get
\begin{eqnarray}
\hat H_0 \: = \: \sum_{{\vec k}, \sigma} \: \left[ \: \epsilon_{{\vec k},
\sigma}^c \: \hat c^{\dagger}_{{\vec k}, \sigma} \: \hat c_{{\vec k},
\sigma} \: + \: \epsilon_{{\vec k}, \sigma}^f \: \hat f^{\dagger}_{{\vec k},
\sigma} \: \hat f_{{\vec k}, \sigma} \: + \: \left( \: V_{\vec k} \:
\hat c^{\dagger}_{{\vec k}, \sigma} \: \hat f_{{\vec k}, \sigma} \:
+ \: h.c. \: \right) \: \right] \: .
\label{ez34}
\end{eqnarray}

The diagonalization of $\hat H_0$ can now be simply done. For this reason
we have to introduce the row vector $W^{\dagger}_{{\vec k}, \sigma} \: = \:
( \: \hat c^{\dagger}_{{\vec k}, \sigma} \: , \: \hat f^{\dagger}_{{\vec k},
\sigma} \: ) \: ,$ and the $( 2 \times 2)$ matrix $\tilde M$ with components
$M_{1, 1} \: = \: \epsilon_{\vec k}^c \: , \: \: M_{1, 2} \: = \: V_{\vec k}
\: , \: \: M_{2, 2} \: = \: \epsilon_{\vec k}^f \: , \: \: 
M_{2, 1} \: = \: V^{*}_{\vec k}$.
The $\hat H_0$ from Eq.(\ref{ez34}) will contain under the sum over 
${\vec k}$ the expression $( \: W_k^{\dagger} \: \tilde M  \: W_k \: )$. 
The diagonalization of $\hat H_0$ in ${\vec k}$-space reduces to the secular 
equation written for the matrix $\tilde M$. We obtain from this
$\left( \: \epsilon_{\vec k}^c \: - \: E_{\vec k} \: \right) \: \left( \:
\epsilon_{\vec k}^f \: - \: E_{\vec k} \: \right) \: - \: | \: V_{\vec k}
\: |^2 \: = \: 0 \: ,$ from where, as expected, two bands arise $( i \: = \:
1, \: 2 )$
\begin{eqnarray}
E_{{\vec k}, i} \: = \: \frac{1}{2} \: \left[ \: \epsilon_{\vec k}^c \: + 
\: \epsilon_{\vec k}^f \: \pm \: T_{\vec k} \: \right] \: .
\label{ez36}
\end{eqnarray}
The expression of $T_{\vec k}$ in Eq.(\ref{ez36}) is given by
$T_{\vec k} \: = \: \sqrt{ \: ( \: \epsilon_{\vec k}^c \: - \: 
\epsilon_{\vec k}^f \: )^2 \: + \: 4 \: | \: V_{\vec k} \: |^2 } \: .$
If we are using now Eqs.(\ref{ez4},\ref{ez401}) in the 
expressions given for $\epsilon_{\vec k}^c, \: \epsilon_{\vec k}^f$,
and $V_{\vec k}$ in terms of the starting Hamiltonian parameters in
Eqs.(\ref{ez26},\ref{ez31}), we realize that the following equality holds
\begin{eqnarray}
| \: V_{\vec k} \: |^2 \: = \: ( \: K \: - \: \epsilon_{\vec k}^c \: ) \: 
( \: K \: - \: \epsilon_{\vec k}^f \: ) \: .
\label{ez38}
\end{eqnarray}
This is a lengthy but straightforward calculation which in fact, easily can be 
done. Using now Eq.(\ref{ez38}) for $T_{\vec k}$, we find
$T^2_{\vec k} \: = \: ( \: \epsilon_{\vec k}^c \: + \: \epsilon_{\vec k}^f \:
- \: 2 \: K)^2 \: ,$ so the band structure given by Eq.(\ref{ez36}) becomes
\begin{eqnarray}
E_{{\vec k}, 1} \: = \: K \: = \: constant \: , \quad
E_{{\vec k}, 2} \: = \: \epsilon_{\vec k}^c \: + \: \epsilon_{\vec k}^f \:
- \: K \: .
\label{ez40}
\end{eqnarray}
From Eq.(\ref{ez4}) we can see that $K \: \geq \: 0$, since $K$ is positive
definite. Furthermore, we have $K \: \geq \: \epsilon_{\vec k}^c$ as well. 
Indeed, using Eqs.(\ref{ez4},\ref{ez26}) we obtain
\begin{eqnarray}
&& K \: - \: \epsilon_{\vec k}^c \: = \: \sum_{i = 1}^4 \: | \: a_{i,c} \:
|^2 \: + \: [ \: ( \: a_{1,c}^{*} \: a_{2,c} \: + \: a^{*}_{4,c} \: a_{3,c} \:
) \: e^{ - \: i \: k_x } \: + \: c.c. \: ] \: + \: [ \: ( \: a_{1,c}^{*} \:
a_{4,c} \: + \: a_{2,c}^{*} \: a_{3,c} \: ) \times
\nonumber\\
&& e^{ - \: i \: k_y } \:
+ \: c.c. \: ] \: + \: [ \: a_{1,c}^{*} \: a_{3,c} \: e^{ - \: i \:
( \: k_x \: + \: k_y \: ) } \: + \: c.c. \: ] \: + \: [ \: a_{2,c}^{*} \:
a_{4,c} \: e^{ - \: i \: ( \: k_y \: - \: k_x \: ) } \: + \: c.c. \: ] \: .
\label{ez40a}
\end{eqnarray}
Introducing now the notations
\begin{eqnarray}
&&a_{1,c}' \: = \: a_{1,c} \: e^{ + \: i \: \frac{k_x \: + \: k_y}{2}} \: , \:
a_{2,c}' \: = \: a_{2,c} \: e^{ - \: i \: \frac{k_x \: - \: k_y}{2}} \: , 
\nonumber\\
&&a_{3,c}' \: = \: a_{3,c} \: e^{ - \: i \: \frac{k_x \: + \: k_y}{2}} \: , \:
a_{4,c}' \: = \: a_{4,c} \: e^{ + \: i \: \frac{k_x \: - \: k_y}{2}} \: ,
\label{ez40b}
\end{eqnarray}
the exponential factors disappear from Eq.(\ref{ez40a}), and for 
$K \: - \: \epsilon_{\vec k}^c$ we find (see also Eq.(\ref{ez52}))
\begin{eqnarray}
&& K \: - \: \epsilon_{\vec k}^c \: = \: ( \: a_{1,c}' \: + \: a_{2,c}' \:
+ \: a_{3,c}' \: + \: a_{4,c}' \: ) \: ( \: a_{1,c}'^{*} \: + \: a_{2,c}'^{*}
\: + \: a_{3,c}'^{*} \: + \: a_{4,c}'^{*} \: ) \: =
\nonumber\\
&& | \: a_{1,c}' \: + \: a_{2,c}' \: + \: a_{3,c}' \: + \: a_{4,c}'
\: |^2 \: .
\label{ez40c}
\end{eqnarray} 
However, $K \: \geq \: \epsilon^c_{\vec k}$ via Eq.(\ref{ez38}) means 
$K \: \geq \: \epsilon_{\vec k}^f$ as well, since $| \: V_{\vec k} \: |^2$ is
a non-negative number. As a consequence, the band structure obtained in 
Eq.(\ref{ez40}) contains an upper band that is completely flat 
$ ( \: E_{{\vec k}, 1} \: = \: E_1 \: = \: K \: ) \: ,$ and a lower, normal, 
${\vec k}$ - dependent band $ ( \: E_{{\vec k}, 2} \: ) $ with dispersion. 
The ${\vec k}$ dependent gap between these two bands is given by
$\Delta_{\vec k} \: = \: E_1 \: - \: E_{{\vec k}, 2} \: = \: 2 \: K \: - \:
( \: \epsilon_{\vec k}^c \: + \: \epsilon_{\vec k}^f \: ) \: ,$
and based on Eqs.(\ref{ez38}, \ref{ez40a}) $\Delta_{\vec k} \: \geq \: 0$ 
holds. Introducing $\Delta \: = \: Min [ \: \Delta_{\vec k} \: ]$, we have 
$\Delta \: \geq \: 0$. Being important below, at this point we mention that 
$\prod_{\vec k} \: \Delta_{\vec k} \: > \: 0$ means $\Delta \: > \: 0$ as 
well. From physical point of view, $\Delta \: \geq \: 0$ means that the 
diagonalized bands from Eq.(\ref{ez40}) are never intersecting. In case of
$\Delta \: = \: 0$ $ , \: E_{{\vec k}, 1}$ and $E_{{\vec k}, 2}$ touch each
other, while for $\Delta \: > \: 0$ the diagonalized bands are completely 
separated.  

From Eq.(\ref{ez40}) it can be observed, that the conditions from 
Eqs.(\ref{ez4},\ref{ez401}) that allows the transformation of the starting 
Hamiltonian Eq.(\ref{ez01}), into $\hat H$ from Eq.(\ref{ez12}), are exactly 
the conditions that give a band structure containing a completely flat (i.e. 
dispersion-less) upper band seen in Eq.(\ref{ez40}).

When the system is interacting and the $U \: > \: 0$ Hubbard term is present,
the Hamiltonian becomes $\hat H$ given in Eq.(\ref{ez11}). In the 
ground-state, because of $\hat P' \: | \: \Psi^U_g \: \rangle \: = \: 0$, 
the effective Hamiltonian has in fact exactly the form of $\hat H_0$ \cite{E1},
excepting that $E_f$ is renormalized as $\tilde E_f \: = \: E_f \: + \: U$ 
(i.e. the condition from Eq.(\ref{ez401}) has to be changed to that given in 
Eq.(\ref{ez401a})), and the energy scale, as seen from Eq.(\ref{ez14}),
is shifted with $U \: N_{\Lambda}$. As a consequence, effectuating the band 
structure calculation as presented above, using instead of $E_f$ the 
$\tilde E_f$ value, we re-obtain (shifted with $U \: N_{\Lambda}$) for the 
ground-state the structure presented in Eq.(\ref{ez40}). The same holds for 
excited states which give $\hat P' \: | \: \Psi \: \rangle \: = \: 0$ as well.

It is interesting to mention at this point, that a two - band
system with a band structure as given in Eq.(\ref{ez40}), above half filling
(i.e. with more than two electrons per lattice site), has a well defined Fermi
energy positioned at $e_F \: = \: E_1 \: = \: K \:$ (where $K$ is a ${\vec k}$ 
independent constant), but the Fermi momentum ${\vec k}_F$ is not definable. 
A such type of system has no Fermi surface in ${\vec k}$-space, so for this 
case, the Luttinger theorem is without meaning.

Another aspect that has to be accentuately underlined, is the fact that the 
Hubbard interaction gives effectively its contribution in the flattening of 
the upper diagonalized band $E_{{\vec k},1}$ in the case in which the here
reported solutions are valid. In order to understand this, first let as
mention that for $U \: = \: 0$, as seen from Eq.(\ref{Ueq0vf}), the solutions
from Eqs.(\ref{ez17},\ref{ez21}) are not applicable. Let turn then to the
$U \: \ne \: 0$ case, and analyze a concrete pedagogical example. Consider for
example the Hamiltonian parameters $t_{c,x} \: = \: t_{c,y} \: = \: 12, \: \:
t_{c,x+y} \: = \: t_{c,y-x} \: = \: - 4, \: \: t_{f,x} \: = \: t_{f,y} \: = \:
3, \: \: t_{f,x+y} \: = \: t_{f,y-x} \: = \: -1, \: \: V_0 \: = \: 18, \: 
V_{1,x} \: = \: V_{1,y} \: = \: -6, \: \: V_{2,x+y} \: = \: V_{2,y-x} \: = \:
2,$ and $E_f \: = \: 2$. The written numbers are expressed in $|t_{f,x+y}|$ 
units. This particular case describe an isotropic situation, with $V^{cf} \:
= \: V^{fc} \: = \: V$, which has been chosen only to be easy for the reader
to follow, without diminishing the general physical content of the behavior
reflected by Eq.(\ref{ez40}). The $E_{{\vec k},1}$ band from 
Eq.(\ref{ez36}) with $U \: = \: 0$ and parameters given above, is presented
in the upper plot of Fig. 2. (the $k_x$ and $k_y$ values cover the first 
Brillouin-zone, i.e. $[- \pi, + \pi]$ in units of the lattice constant). 
As can be seen, for this non-interacting case we have a normal ${\vec k}$ 
dependent band, the system is metallic, and the system of equations
Eqs.(\ref{ez4},\ref{ez401}) has no solutions (i.e. at $U = 0$ we are situated
outside of ${\cal{P}}_{\hat H}$). Hovewer, turning the interaction
on (note that for $U \: \ne \: 0$ we have $\tilde E_f$ instead of $E_f$ in 
Eq.(\ref{ez36}), and Eq.(\ref{ez401}) has to be changed with 
Eq.(\ref{ez401a})), the presence of the interaction starts to flatten the 
$E_{{\vec k},1}$ band. Indeed, as seen from Fig. 2.
(middle plot: $U \: = \: 10$, bottom: $ U \: = \: 20$), the $E_{{\vec k}, 1}$
surface starts to flatten with increasing $U$ values. The completely flat
$E_{{\vec k},1}$ case given in Eq.(\ref{ez40}) is obtained for $U \: = \: 25$,
which at the level of the system of equations Eqs.(\ref{ez4},\ref{ez401a}), is
represented by the solution $a_{1,c} \: = \: 3 + \sqrt{5}, \: \: a_{2,c} \: =
\: -2, \: \: a_{3,c} \: = \: 3 - \sqrt{5}, \: \: a_{4,c} \: = \: -2, \: \:
a_{1,f} \: = - (\sqrt{5} + 3)/2, \: \: a_{2,f} \: = \: 1, \: \: a_{3,f} \: = 
\: (\sqrt{5} - 3)/2, \: \: a_{4,f} \: = \: 1.$ When this solution emerges
at $U = 25$, the Hubbard interaction has pushed the system inside 
${\cal{P}}_{\hat H}$, so its role is fully active in obtaining the interacting
ground-states described here.     

\subsection{Decomposition into composite operators}
 
We have to mention that obtaining the diagonalized band picture for $\hat H$
presented in Eq.(\ref{ez40}), the mathematical description can be 
given in ${\vec k}$-space in term of new (composite) fermionic operators,
which creates composite fermions in the upper and lower band. To see this,
we note that the following relation holds 
\begin{eqnarray}
&&\epsilon^c_{\vec k} \: \hat c^{\dagger}_{{\vec k}, \sigma} \: 
\hat c_{{\vec k}, \sigma} \: + \: \epsilon^f_{\vec k} \: 
\hat f^{\dagger}_{{\vec k}, \sigma} \: \hat f_{{\vec k}, \sigma} \: + \: 
\left( \: | \: V_{\bf k} \: | \: e^{ i \: \phi_V } \: \hat c^{\dagger}_{
{\vec k}, \sigma} \: f_{{\vec k}, \sigma} \: + \: h. c. \: \right) \: =
\nonumber\\
&& K \: {\hat C}^{\dagger}_{{\vec k}, 1, \sigma} \: {\hat C}_{{\vec k}, 1,
\sigma} \: + \: ( \: \epsilon^c_{{\vec k}, \sigma} \: + \: \epsilon^f_{
{\vec k}, \sigma} \: - \: K \: ) \: {\hat C}^{\dagger}_{{\vec k}, 2, \sigma}
\: {\hat C}_{{\vec k}, 2, \sigma} \: ,
\label{ez41a}
\end{eqnarray}
where for $j \: = \: 1, 2$ the operators ${\hat C}_{{\vec k}, j, \sigma}$ 
are defined as
\begin{eqnarray}
{\hat C}_{{\vec k}, j, \sigma} \: = \: \frac{ ( - 1)^j}{ \sqrt{
\Delta_{\bf k} } } \: \left( \: \sqrt{ K \: - \: \epsilon_{\vec k}^c } \:
{\hat f}_{{\vec k}, \sigma} \: - \: ( - 1 )^j \: e^{+ \: (-1)^j \: i \:
\phi_V} \: \sqrt{ K \: - \: \epsilon_{\vec k}^f } \: {\hat c}_{{\vec k},
\sigma} \: \right) \: e^{ + \: i \: \theta_j} \: .
\label{ez41b}
\end{eqnarray}
From Eq.(\ref{ez41a}) it can be seen that $\phi_V$ represents the argumentum 
of $V_{\bf k}$, i.e. $ V_{\bf k} \: = \: \mid \: V_{\bf k} \: \mid \: 
e^{ i \: \phi_V} $. Furthermore, $\theta_1$ and $\theta_2$ represent two 
arbitrary, independent, and ${\vec k}$ independent phases. Using 
Eqs.(\ref{ez40},\ref{ez41a}), the operator $\hat H_0$ can be written now in 
a simple form
\begin{eqnarray}
{\hat H}_0 \: = \: \sum_{k, \sigma} \: \left( \: E_{{\vec k}, 1} \: 
{\hat C}_{{\vec k}, 1, \sigma}^{\dagger} \: {\hat C}_{{\vec k}, 1, \sigma} \:
+ \: E_{{\vec k}, 2} \: {\hat C}_{{\vec k}, 2, \sigma}^{\dagger} \: 
{\hat C}_{{\vec k}, 2, \sigma} \: \right) \: . 
\label{ez41c}
\end{eqnarray}

The expression presented in Eq.(\ref{ez41c}) shows that the 
operators defined in Eq.(\ref{ez41b}) annihilate (their adjoint create) 
particles from (into) the ,,diagonalized bands'' $j \: = \: 1$ (upper flat 
band), and $j \: = \: 2$ (lower band with dispersion), respectively. Since
$\left\{ \: \hat C_{{\vec k}, j, \sigma} \: , \: \hat C^{\dagger}_{{\vec k}',
 j', \sigma'} \: \right\} \: =  \: \delta_{{\vec k}, {\vec k}'} \:
\delta_{j, j'} \: \delta_{\sigma, \sigma'} \: , \quad
\left\{ \: \hat C_{{\vec k}, j, \sigma} \: , \: \hat C_{{\vec k}', j',
\sigma'} \: \right\} \: = \: \left\{ \: \hat C^{\dagger}_{{\vec k}, j, 
\sigma} \: , \: \hat C^{\dagger}_{{\vec k}', j', \sigma'} \: 
\right\} \: =  \: 0 \: ,$ the $\hat C_{{\vec k}, j, \sigma}$ operators 
represent new, rigorous, canonical and anti-commuting Fermionic operators.

The transformation into Eq.(\ref{ez41c}) physically is more deep as seems to 
be at first view. In order to see this, we have to analyze the Fourier 
transform of the plaquette operators $\hat A_{I, \sigma}$. Starting from 
Eq.(\ref{ez1}), the Fourier transform of $\hat A^{\dagger}_{I, \sigma}$ can 
be written as
\begin{eqnarray}
\hat A^{\dagger}_{I, \sigma} \: = \: \sum_{\vec k} \: e^{ i \: {\vec k} 
\cdot {\vec r} } \: [ \: X_c( {\vec k} ) \: \hat c^{\dagger}_{{\vec k},
\sigma} \: + \: X_f( {\vec k} ) \: \hat f^{\dagger}_{{\vec k}, \sigma} 
\: ] \: ,
\label{ez47}
\end{eqnarray}
where, for $b \: = \: c, f$ we have
\begin{eqnarray}
X_b( {\vec k} ) \: = \: a^{*}_{1,b} \: + \: a^{*}_{2,b} \: e^{ i \: {\vec k} 
\cdot {\vec x}} \: + \: a^{*}_{3,b} \: e^{ i \: {\vec k} \cdot ( {\vec x} \:
+ \: {\vec y} ) } \: + \: a^{*}_{4,b} \: e^{ i \: {\vec k} \cdot {\vec y} } 
\: .
\label{ez48}
\end{eqnarray}
Effectuating the product over the ordered plaquette index $I$ in 
$\prod_{I} \: \hat A^{\dagger}_{I, \sigma}$  (such a product emerges in the 
ground-state wave function from Eqs.(\ref{ez17},\ref{ez21})), given by the 
anti-commutation rules of the starting $\hat f, \: \hat c$ operators, only a 
product containing different ${\vec k}$ indices survives. Introducing the 
notation
$Y \: = \: \sum_{\bar P} \: ( - 1 )^{\bar p} \: exp [ \: i \: ( \: {\vec r}_1
\cdot {\vec k}_{i1} \: + \: {\vec r}_2 \cdot {\vec k}_{i2} \: + \: ... \: + 
\: {\vec r}_{N_{\Lambda} } \cdot {\vec k}_{i N_{\Lambda}}\: ) \: ] \: ,$
where $\sum_{\bar P}$ denotes a sum over all possible permutations of
$(1, \: 2, ..., \: N_{\Lambda})$ to $(i_1, \: i_2, ... , \: i_{N_{\Lambda}}
)$, and $\bar p$ represents the number of pair permutations in a given 
$\bar P$, we obtain
\begin{eqnarray}
\prod_{I} \: \hat A^{\dagger}_{I, \sigma} \: = \: Y \: \prod_{\vec k} \:
[ \: X_c( {\vec k} ) \: \hat c^{\dagger}_{{\vec k}, \sigma} \: + \:
X_f( {\vec k} ) \: \hat f^{\dagger}_{{\vec k}, \sigma} \: ] \: .
\label{ez50}
\end{eqnarray}
As can be seen from Eq.(\ref{ez50}), the contribution into the norm of the
ground state wave function given by $\prod_{I} \: \hat A^{\dagger}_{I,
\sigma}$ is proportional to ${\cal R} \: = \: \prod_{\vec k} \: {\cal R}_{
\vec k}$ where $ {\cal R}_{\vec k} \: = \: [ \: | \: X_c( {\vec k} ) \: |^2 
\: + \: | \: X_f( {\vec k} ) \: |^2 \: ] \: .$
However, using Eq.(\ref{ez26}) together with the definitions given in Sec.III.
for $\epsilon^b_{\vec k}$ and Eqs.(\ref{ez4},\ref{ez48}), we find
\begin{eqnarray}
| \: X_c( {\vec k} ) \: |^2 \: = \: K \: - \: \epsilon_{\vec k}^c \: , \quad
| \: X_f( {\vec k} ) \: |^2 \: = \: K \: - \: \epsilon_{\vec k}^f \: , \quad
X_c( {\vec k} ) \: X_f^{*}( {\vec k} ) \: = \: - \: V_{\vec k} \: .
\label{ez52}
\end{eqnarray}
For example, in the case of the first relation from Eq.(\ref{ez52}), we
simply must use $X^{*}_c( {\vec k} )$ from Eq.(\ref{ez48}) and multiply it
with $exp [ \: ( i / 2 ) \: ( \: k_x \: + \: k_y \: ) \: ]$ in order to see 
the expression of $( \: K \: - \: \epsilon^c_{\vec k} \: )$ presented in 
Eq.(\ref{ez40c}). The remaining two equalities can be similarly deduced.

From the first two relations of Eq.(\ref{ez52}) we observe that
the contribution of the operator $\hat F^{(1)}$ (or $\hat F^{(2)}$)
into the norm of the ground state wave function  is determined via 
Eq.(\ref{ez50}) by
\begin{eqnarray}
{\cal R}_{\vec k} \: = \: \Delta_{\vec k} \: = \: | \: X_c( {\vec k} ) \: |^2
\: + \: | \: X_f( {\vec k} ) \: |^2 \: = \: 2 \: K \: - \: ( \: 
\epsilon_{\vec k}^c \: + \: \epsilon_{\vec k}^f \: ) \: .
\label{ez53}
\end{eqnarray}

Let us now introduce instead of the initial plaquette operators
$\hat A_{I, \sigma}$, new plaquette operators ,,normalized to unity'', i.e.
whose contribution under the $\prod_I$ product give unity into
the norm of the wave function. In this case instead of $\hat A_{I, \sigma}$
we must use  
\begin{eqnarray}
\hat B_{I,\sigma} \: = \: \left( \: \frac{ | \: Y \: |^{-1} }{
\sqrt{ \prod_{\vec k} \: \Delta_{\vec k} } } \: \right)^{\frac{1}{N_{
\Lambda} } } \: \: \hat A_{I, \sigma} \: .
\label{ez200}
\end{eqnarray}
In this case we have instead of $\prod_{I} \: {\hat A}^{\dagger}_{I, \sigma}$,
the product 
\begin{eqnarray}
\prod_I \: \hat B^{\dagger}_{I, \sigma} \: = \: \prod_{\vec k} \: \left[ \:
\frac{1}{ \sqrt{ \Delta_{\vec k} } } \: \left( \: X_c( {\vec k} ) \:
\hat c^{\dagger}_{{\vec k}, \sigma} \: + \: X_f( {\vec k} ) \:
\hat f^{\dagger}_{{\vec k}, \sigma} \: \right) \: \right] \: = \:
\prod_{\vec k} \: \hat B^{\dagger}_{{\vec k}, 2, \sigma} \: ,
\label{ez201}
\end{eqnarray}
where
\begin{eqnarray}
\hat B^{\dagger}_{{\vec k}, 2, \sigma} \: = \: \frac{1}{ \sqrt{
\Delta_{\vec k} } } \: \left( \: X_c( {\vec k} ) \: \hat c^{\dagger}_{
{\vec k}, \sigma} \: + \: X_f( {\vec k} ) \: \hat f^{\dagger}_{{\vec k},
\sigma} \: \right) \: .
\label{ez202}
\end{eqnarray}
Now let us denote $X_c( {\vec k} ) = | X_c( {\vec k} ) | \: e^{ i \phi_c}$, 
$X_f( {\vec k} ) = | X_f( {\vec k} ) | \: e^{ i \phi_f} $, and 
$V_{\vec k} = | V_{\vec k} | \: e^{ i \phi_V}$. 
Using these notations, from Eq.(\ref{ez41b}) and first two equations from 
Eq.(\ref{ez52}) we find
$\hat C^{\dagger}_{{\vec k}, 2, \sigma} \: = \: \Delta_{\vec k}^{-1/2} \:
( \: | \: X_c( {\vec k} ) \: | \: \hat c^{\dagger}_{{\vec k}, \sigma}
\: - \: e^{ - \: i \: \phi_V} \: | \: X_f( {\vec k} ) \: | \: 
\hat f^{\dagger}_{{\vec k}, \sigma} \: ) \: .$
From Eq.(\ref{ez202}) we have however
$\hat B^{\dagger}_{{\vec k}, 2, \sigma} \: = \: \Delta_{\vec k}^{-1/2} \:
( \: | \: X_c( {\vec k} ) \: | \: e^{ i \: \phi_c} \: \hat c^{\dagger}_{
{\vec k}, \sigma} \: + \: | \: X_f( {\vec k} ) \: | \: e^{ i \: \phi_f} \:
\hat f^{\dagger}_{{\vec k}, \sigma} \: ) \: = \: e^{ i \: \phi_c} \:
\Delta_{\vec k}^{-1/2} \: ( \: | \: X_c( {\vec k} ) \: | \: 
\hat c^{\dagger}_{{\vec k}, \sigma} \: + \: | \: X_f( {\vec k} ) \: | \:
 e^{ i \: ( \: \phi_f \: - \: \phi_c \: ) } \: \hat f^{\dagger}_{{\vec k},
\sigma} \: ) \: .$ From the third relation of Eq.(\ref{ez52}) via 
Eqs.(\ref{ez38},\ref{ez52}) we obtain $e^{ i \: ( \: \phi_c \: - \: \phi_f \:
) } \: = \: - \: e^{ i \: \phi_V}$, which means $e^{ i \: ( \: \phi_f \:
- \: \phi_c \: ) } \: = \: - \: e^{ - \: i \: \phi_V  } \: , $ and the 
expression of $\hat B^{\dagger}_{{\vec k}, 2, \sigma}$ becomes
\begin{eqnarray}
\hat B^{\dagger}_{{\vec k}, 2, \sigma} \: = \: e^{ i \: \phi_c} \:
\hat C^{\dagger}_{{\vec k}, 2, \sigma} \: .
\label{ez208}
\end{eqnarray}
This means that the Fourier transform of $\hat B_{I, \sigma}$ (i.e. the
normalized $\hat A_{I, \sigma}$), is in fact $\hat C_{{\vec k}, 2, \sigma}$ 
if we fix the phase $\theta_2$ from Eq.(\ref{ez41b}) to $\theta_2 \: = \:
- \: \phi_c$.

Similar to $\hat B^{\dagger}_{{\vec k}, 2, \sigma}$, we can introduce
the normalized $\hat B^{\dagger}_{{\vec k}, 1, \sigma} \: = \:
e^{ - \: i \: \phi_c} \: \hat C^{\dagger}_{{\vec k}, 1, \sigma}$, which
creates a particle in the upper band, by fixing the $\theta_1$ phase as 
$\theta_1 \: = \: \phi_c$. We obtain
\begin{eqnarray}
\hat B^{\dagger}_{{\vec k}, 1, \sigma} \: = \: \frac{1}{\sqrt{
\Delta_{\vec k} } } \: \left( \: X_f^*( {\vec k} ) \: \hat c^{\dagger}_{
{\vec k}, \sigma} \: - \: X_c^*( {\vec k} ) \: \hat f^{\dagger}_{{\vec k},
\sigma} \: \right) \: .
\label{ez208a}
\end{eqnarray}
Since in comparison with $\hat C_{{\vec k}, j, \sigma}$, only a phase factor
difference emerges, the canonical anti-commutation relations remain true for 
$\hat B_{{\vec k}, j, \sigma}$ ($j \: = \: 1, 2$) operators as well. Using
Eqs.(\ref{ez208},\ref{ez208a}), the initial operators can be also expressed
via
\begin{eqnarray}
\hat f_{{\vec k}, \sigma} \: = \: \frac{ X_f( {\vec k} ) \: \hat B_{{\vec k},
2, \sigma} \: - \: X_c^*( {\vec k} ) \: \hat B_{{\vec k}, 1, \sigma} }{\sqrt{
\Delta_{\vec k} } } \: , \quad \hat c_{{\vec k}, \sigma} \: = \: \frac{
X_c( {\vec k} ) \: \hat B_{{\vec k}, 2, \sigma} \: + \: X_f^*( {\vec k} ) \:
\hat B_{{\vec k}, 1, \sigma} } {\sqrt{ \Delta_{\vec k} } } \: .
\label{ez208b}
\end{eqnarray}
With the composite operators introduced, for the $\hat H_0$ operator we obtain
\begin{eqnarray}
\hat H_0 \: = \: \sum_{j = 1,2} \: \sum_{{\vec k}, \sigma} \: E_{{\vec k},
j, \sigma} \: \hat B^{\dagger}_{{\vec k}, j, \sigma} \:
\hat B_{{\vec k}, j, \sigma} \: ,
\label{ez208c}
\end{eqnarray}
and the ground-state wave functions presented in Sec.II. becomes
\begin{eqnarray}
&&| \: \Psi^U_g \: \rangle \: = \: \left[ \: \prod_{\vec k} \: \left( \:
\hat B^{\dagger}_{{\vec k}, 2, \uparrow} \: \hat B^{\dagger}_{{\vec k}, 2,
\downarrow} \: \right) \: \right] \: \hat F^{(3)} \: | \: 0 \: \rangle \: , 
\nonumber\\
&&| \: \Psi^U_{g,d} \: \rangle \: = \: \left[ \: \prod_{\vec k} \: \left(
\: \hat B^{\dagger}_{{\vec k}, 2, \uparrow} \: \hat B^{\dagger}_{{\vec k},
2, \downarrow} \: \right) \: \right] \: \hat F^{(3)} \: \hat F^{(4)} \:
| \: 0 \: \rangle \: ,
\label{ez203}
\end{eqnarray}
where the ordered product over ${\vec k}$ has to be taken over the whole first
Brillouin zone. The norm of the ground-state wave functions from 
Eq.(\ref{ez203}) is entirely determined by $\hat F^{(\beta)}$ with 
$\beta \: = \: 3, 4$.  

We however stress, that the ${\vec k}$ representations for the deduced 
ground-state wave functions presented in Eqs.(\ref{ez203}) are valid only if
the initial ground-state wave functions from Eqs.(\ref{ez17},\ref{ez21}) have
nonzero norm. This is the case if $\prod_{\vec k} \: \Delta_{\vec k} \:
\ne \: 0$ holds (see Eq.(\ref{ez200})), i.e. the diagonalized bands 
$E_{{\vec k}, j =  1,2}$ are completely separated. In order to find such 
situations, mathematically is sufficient to have for all ${\vec k}$ values
from the first Brillouin zone $| X_c ( {\vec k} ) | \: > \: 0$. 

\section{Magnetic properties}

Continuing the study of physical properties of the deduced ground-state,
the following natural step is to analyze its magnetic properties. This is 
motivated by the existence of ferro-magnetism in some flat-band 
models\cite{magnet}. Because of this reason, we have to check if something
similar is present for our solutions, or not. Starting from this motivation,
we obtain however paramagnetic properties (i.e. large spin degeneracy) for 
the deduced ground states. The reason for this is the following one.

As can be seen from Eq.(\ref{ez16a}), the obtained ground-state wave functions
have a large degeneracy. With this observation in mind, let us consider first
the $N \: = \: 3 \: N_{\Lambda}$ (i.e $3/4$ filling) case. We have a $d_g \: 
= \: 2^{N_{\Lambda}}$ fold degeneracy in $| \: \Psi^U_g \: \rangle$ due to 
the up or down orientation possibilities of the $N/3 \: = \: N_{\Lambda}$ 
electrons from the system, since $2 \: N_{\Lambda}$ particles fill up 
completely the lower band, so their contribution in the total spin is zero. 
From the point of view of the $\alpha_{{\bf i}, \sigma}$ coefficients 
entering in $\hat F^{(3)}$ as given in Eq.(\ref{ez16}), this means that it 
is possible for us to obtain $d_g$ linearly independent contributions in 
$| \: \Psi^U_g \: \rangle$, by choosing arbitrary $\alpha_{{\bf i}, 
\uparrow}$ and $\alpha_{{\bf i}, \downarrow}$ coefficients. The $S^z$ value 
of these states (S being the total spin) is situated between $ [ \: - \:
N_{\Lambda} / 2 \: , \: + \: N_{\Lambda} / 2 \: ]$. The contributions from 
these states with strictly $S^z \: = \: N_{\Lambda} / 2$ can be obtained by 
choosing $\alpha_{{\bf i}, \uparrow} \: = \: 1$ and 
$\alpha_{{\bf i}, \downarrow} \: = \: 0$ for all ${\bf i}$ sites. This 
maximal $S^z$ for the system can be achieved in an unique way, therefore 
there is an unique state (apart from the trivial $( \: 2 \: S \: + \: 1 \: )$
fold degeneracy) among the linearly independent contributions in the 
ground-state, with $S \: = \: N_{\Lambda}/2$.

After finding this contribution in $| \: \Psi^U_g \: \rangle$ characterized
by $S \: = \: N_{\Lambda}/2$ total spin, let us chose one single lattice site 
${\bf i}_1$ with $\alpha_{{\bf i}_1, \uparrow} \: = \: 0$ and 
$\alpha_{{\bf i}_1, \downarrow} \: = \: 1$, the remaining sites 
${\bf i} \ne {\bf i}_1$ being maintained with $\alpha_{{\bf i}, \uparrow} \:
= \: 1, \: \alpha_{{\bf i}, \downarrow} \: = \: 0$. Moving the position of 
the ${\bf i}_1$ site along the lattice, we obtain all contributions in 
the ground-state with $S^z \: = \: N_{\Lambda} / 2 \: - \: 1$. Let us denote 
these states by $| \: \Psi_{\beta} \: \rangle$, where $\beta \: = \: 1, 2, 
... N_{\Lambda}$. These states are linearly independent (see Appendix), 
and they span an $N_{\Lambda}$ dimensional subspace since a single particle 
spin can be flipped down independently on $N_{\Lambda}$ lattice sites. 
The mentioned $| \: \Psi_{\beta} \: \rangle$ states are also components of 
$| \: \Psi^U_g \: \rangle$. Since between the $| \: \Psi_{\beta} \: \rangle$ 
wave vectors it must be one term corresponding to the 
$( \: S^z \: = \: N_{\Lambda}/2 \: - \: 1, \: S \: = \: N_{\Lambda}/2 \: )$
state, we have $N_{\Lambda} \; - \: 1$ linearly independent states with 
$S \: = \: N_{\Lambda}/2 \: - \: 1$. As a consequence,
$| \: \Psi^U_g \: \rangle$ contains not only components with $S \: = \:
N_{\Lambda} / 2$, but also components with $S \: = \: N_{\Lambda} / 2 \:
- \: 1$. Continuing this procedure, one can see that every spin value $S$ is 
present among the ground state contributions $N ! \: ( \: 2 \: S \: + 1 \: )
\: / \: [ \: ( \: N/2 \: + \: S \: + \: 1 \: ) ! \: ( \: N/2 \: - \: S \:) !
\: ]$ times (and certainly everyone has $2 \: S \: + \: 1$ components with
different $S^z$ values). The $S \: = \: 0$ subspace has the greatest 
degeneracy, i.e. it has the greatest statistical weight. Therefore, in the 
thermodynamic limit the expectation value of the spin goes to zero, i.e.
the system is paramagnetic. 

In fact, since states with different $S$ can be obtained from 
$| \: \Psi^U_g \: \rangle$ simply by modifying the $\alpha_{{\bf i}, \sigma}$
values in Eq.(\ref{ez16}), it means that calculating the ground-state 
expectation value $\langle \: {\hat S}^2 \: \rangle$ based on
Eqs.(\ref{ez16},\ref{ez17}), we obtain an expression that depends on the
arbitrary coefficients $\alpha_{{\bf i}, \sigma}$. This means that for 
$| \: \Psi^U_g \: \rangle$, the value of the total spin is not fixed, it
possesses a large spin degeneracy, i.e. is paramagnetic.
This property will not be changed even if we take into consideration the
$\hat F^{(4)}$ operator in the ground-state wave function (i.e. doped system),
because we add in this case in the expression of $\langle {\hat S}^2 \rangle$,
contributions depending on $\alpha_{{\bf i}, \sigma}$ multiplied by arbitrary 
constants $C_{\{ {\bf i} \} }$ present in Eq.(\ref{ez20}). 
As a consequence, the detected solutions describes paramagnetic phases.

Before closing this section, we have to mention that large spin degeneracy 
characteristics of the ground-states for strongly correlated systems have
been also reported elsewhere. An interesting result on this line, is that
described by Arita and Aoka \cite{E2}. These authors have found for $t-t'$ 
type Hubbard models ground-states holding simultaneously total spin $S \: = \:
0$ and $S \: = \: S_{max}$ values in the thermodynamical limit. In their
case, the singlet $S \: = \: 0$ component is created by a spiral spin state 
with a spin correlation length as large as the system size, which accompanies
the fully polarized ferromagnetic state with total spin $S_{max}$. We have to
underline, that our case differs semnificatively from that presented in Ref.
\cite{E2}. In contrast to Ref.\cite{E2}, in the ground states from 
Eqs.(\ref{ez17},\ref{ez21}), all total spin $S$ values are present. As
will be exemplified later on (see for example a concrete solution 
described in Eq.(\ref{ins2})), given by the arbitrary nature of the 
coefficients $\alpha_{{\bf i},\sigma}$, a local spin periodicity presence
in the ground-state is rather accidental, instead of a characteristic
(or general) property.

\section{The insulating phase}

Besides the fact that the states we obtained are non-magnetic, their physical
properties still remain to be clarified. These properties depend in fact from
the possible solutions of Eqs.(\ref{ez4},\ref{ez401a}). To have an insight
in these possibilities, let us analyze Eq.(\ref{ez4}). In this study we have 
to consider the coupling constants from the Hamiltonian $\hat H$ known 
variables, and to try to deduce based on them the $a_{i,b}$ parameters 
entering into $\hat A_{I,\sigma}$ presented in Eq.(\ref{ez1}). We may start 
solving this problem by introducing the parameters
\begin{eqnarray}
&& p_1 \: = \: \frac{V^{cf}_{2,x+y}}{t_{f,x+y}} \: = \: 
\frac{a^*_{1,c}}{a^*_{1,f}} \: , \quad
p_2 \: = \: \frac{V^{fc}_{2,x+y}}{t_{f,x+y}} \: = \: 
\frac{a_{3,c}}{a_{3,f}} \: ,
\nonumber\\
&& q_1 \: = \: \frac{V^{cf}_{2,y-x}}{t_{f,y-x}} \: = \:
\frac{a^*_{2,c}}{a^*_{2,f}} \: , \quad 
q_2 \: = \: \frac{V^{fc}_{2,y-x}}{t_{f,y-x}} \: = \:
\frac{a_{4,c}}{a_{4,f}} \: .
\label{ez42}
\end{eqnarray}
Comparing Eqs.(\ref{ez1},\ref{ez42}), we can rewrite the plaquette operator 
$\hat A_{I, \sigma}$ as follows
\begin{eqnarray}
&& \hat A_{I,\sigma} \: =  \: a_{1,f} \: ( \: \hat f_{1,\sigma} \: + \:
p_1^* \: \hat c_{1,\sigma} \: ) \: + \: a_{2,f} \: ( \: \hat f_{2,\sigma} \:
+ \: q_1^* \: \hat c_{2,\sigma} \: ) \: +
\nonumber\\
&& a_{3,f} \: ( \: \hat f_{3,\sigma} \: + \: p_2 \: \hat c_{3,\sigma} \:
) \: + \: a_{4,f} \: ( \: \hat f_{4,\sigma} \: + \: q_2 \: \hat c_{4,\sigma}
\: ) \: .
\label{ez43}
\end{eqnarray}

The physical nature of the ground-state is strongly influenced by the particle
distribution created within the lattice by $| \: \Psi^U_g \: \rangle$ or 
$| \: \Psi^U_{g,d} \: \rangle$. Since $\hat F^{(3)}$ introduces one electron 
on every lattice site, this particle distribution is essentially determined
by the product $\hat F^{(1)} \: \hat F^{(2)}$. But on a given lattice site we 
independently can introduce two electrons with opposite spin, reason for which
we focus on the behavior of $\hat F^{(1)}$. In order to have an image about 
these aspects, we are interested to analyze the situation in which the 
operator $\hat F^{(1)} \: = \: \prod_I \: \hat A^{\dagger}_{I, \sigma}$ 
introduces two electrons (one $c$ and one $f$), at least on a lattice site. 
If a such situation emerges, $\hat F^{(1)}$ creates a non-uniform particle
distribution within the lattice. This is because it introduces in the system 
$N_{\Lambda}$ particles on $N_{\Lambda}$ lattice sites creating somewhere a
double occupancy, which must be followed by an empty site. Using 
Eq.(\ref{ez43}) and multiplying two arbitrary neighboring plaquettes, for a 
site taking place in both plaquettes the following six type of nonzero 
products may emerge 
\begin{eqnarray}
&&I_1 \: = \: ( \hat f^{\dagger}_{i,\sigma} \: + \: p_1 \: 
\hat c^{\dagger}_{i,\sigma} ) \cdot ( \hat f^{\dagger}_{i,\sigma} \: + \:
q_1 \: \hat c^{\dagger}_{i,\sigma} ), \: \:   
I_2 \: = \: ( \hat f^{\dagger}_{i,\sigma} \: + \: p_1 \: 
\hat c^{\dagger}_{i,\sigma} ) \cdot ( \hat f^{\dagger}_{i,\sigma} \: + \:
p_2^* \: \hat c^{\dagger}_{i,\sigma} ) \: ,
\nonumber\\   
&&I_3 \: = \: ( \hat f^{\dagger}_{i,\sigma} \: + \: p_1 \: 
\hat c^{\dagger}_{i,\sigma} ) \cdot ( \hat f^{\dagger}_{i,\sigma} \: + \:
q_2^* \: \hat c^{\dagger}_{i,\sigma} ), \: \:
I_4 \: = \: ( \hat f^{\dagger}_{i,\sigma} \: + \: q_1 \: 
\hat c^{\dagger}_{i,\sigma} ) \cdot ( \hat f^{\dagger}_{i,\sigma} \: + \:
p_2^* \: \hat c^{\dagger}_{i,\sigma} ) \: ,
\nonumber\\
&&I_5 \: = \: ( \hat f^{\dagger}_{i,\sigma} \: + \: q_1 \: 
\hat c^{\dagger}_{i,\sigma} ) \cdot ( \hat f^{\dagger}_{i,\sigma} \: + \:
q_2^* \: \hat c^{\dagger}_{i,\sigma} ), \: \:
I_6 \: = \: ( \hat f^{\dagger}_{i,\sigma} \: + \: p_2^* \: 
\hat c^{\dagger}_{i,\sigma} ) \cdot ( \hat f^{\dagger}_{i,\sigma} \: + \:
q_2^* \: \hat c^{\dagger}_{i,\sigma} ) \: .        
\label{ez44}
\end{eqnarray}
If at least one $I_j$ term from Eq.(\ref{ez44}) is nonzero, we identified
at least a site ${\bf i}$, where $\prod_I \: \hat A^{\dagger}_{I, \sigma}$ 
introduces two particles. Effectuating the products in Eq.(\ref{ez44}) and 
taking into consideration the anti-commutation rules for the fermionic 
creation operators, for $I_j$ from Eq.(\ref{ez44}) we find
\begin{eqnarray}
&&
I_1 \: = \: \hat c^{\dagger}_{i,\sigma} \: \hat f^{\dagger}_{i,\sigma} \:
( \: p_1 \: - \: q_1 \: ), \: \:
I_2 \: = \: \hat c^{\dagger}_{i,\sigma} \: \hat f^{\dagger}_{i,\sigma} \:
( \: p_1 \: - \: p_2^* \: ), \: \:
I_3 \: = \: \hat c^{\dagger}_{i,\sigma} \: \hat f^{\dagger}_{i,\sigma} \:
( \: p_1 \: - \: q_2^* \: ) \: , 
\nonumber\\
&&
I_4 \: = \: \hat c^{\dagger}_{i,\sigma} \: \hat f^{\dagger}_{i,\sigma} \:
( \: q_1 \: - \: p_2^* \: ), \: \:
I_5 \: = \: \hat c^{\dagger}_{i,\sigma} \: \hat f^{\dagger}_{i,\sigma} \:
( \: q_1 \: - \: q_2^* \: ), \: \:
I_6 \: = \: \hat c^{\dagger}_{i,\sigma} \: \hat f^{\dagger}_{i,\sigma} \:
( \: p_2^* \: - \: q_2^* \: ) \: .
\label{ez45}
\end{eqnarray}
Usually for the hybridization matrix elements we have 
$V^{cf}_{\chi} \: = \: V^{fc}_{\chi} \: = \: V_{\chi}$ for an arbitrary
element $\chi$, reason for which we focus in the remaining part of the paper
to this case (the results presented up to the end of Sec.IV. remaining
valid for the general $V_{\chi}^{cf} \: \ne \: V_{\chi}^{fc}$ situation as 
well). This means that we have in fact
\begin{eqnarray}
p \: = \: p_1 \: = \: p_2 \: , \quad q \: = \: q_1 \: = \: q_2 \: .
\label{ez46}
\end{eqnarray}
Now if we consider only real coupling constants in the Hamiltonian $\hat H$,
in Eq.(\ref{ez45}) only the terms containing $\delta \: = \: ( \: p \: -
\: q \: )$ survives. From physical point of view, if $\delta \: = \: 0$, the 
wave function $| \: \Psi^U_g \: \rangle$ introduces on every lattice site 
the same number of electrons. Since hopping matrix elements for a such type 
of state are all zero, the ground-state is completely localized, i.e. 
represents a Mott-insulator.

After this step, considering $p$ and $q$ pure real variables (i.e. all
coupling constants from $\hat H$ are pure real), we have to check if 
the solutions allowed by Eqs.(\ref{ez4},\ref{ez401a}) gives $\delta \: = \: 
0$ or not. For this reason two cases have to be enumerated. 

\subsection{The symmetric case}

In the first step we analyze the $p \: = \: q$ situation, called symmetric 
case. Physically a such type of parameter region can be obtained, for example,
when the lattice has a non-distorted unit cell. In these circumstances 
$t_{b, x+y} \: = \: t_{b, y-x}$ and $V_{2,x+y} \: = \: V_{2,y-x}$,
i.e. $(p-q) \: = \: 0$, so all $I_j \: = \: 0$ from Eq.(\ref{ez45}). 
(We underline that rigorously the condition $p \: = \: q$ means 
$V_{2,x+y} \: V_{2,y-x} \: = \: t_{f,x+y} \: t_{c,y-x}$, which exceeds in 
fact the non-distorted unit cell situation.). We mention here, that for 
$p \: = \: q$ the system of equations from Eq.(\ref{ez4}) admits only 
solutions of the type $p \: = \: p^*$. In this case 
$\hat A^{\dagger}_{I,\sigma} \: = \: \sum_{i=1}^{4} \: a_{i,f}^{*} \: ( \:
p \: \hat c^{\dagger}_{i,\sigma} \: + \: \hat f^{\dagger}_{i,\sigma} \: )$, 
and the product $\prod_I \: \hat A^{\dagger}_{I,\sigma}$ from the 
ground-state wave function can be replaced by $ \tilde F_{\sigma} \: = \:
\prod_{\bf i} \: \tilde F_{{\bf i},\sigma}$, where $\tilde F_{{\bf i},\sigma} 
\: = \: ( \: p \: c^{\dagger}_{{\bf i},\sigma} \: + \: \hat f^{\dagger}_{
{\bf i}, \sigma} \: )$. We note, that the product in $\tilde F_{\sigma}$  
is a product over lattice sites instead of plaquettes. The contribution into
the norm of $\tilde F_{\sigma}$ is well defined and equal with
$( \: 1 \: + \: p^2 \: )^{N_{\Lambda}}$. The ground-state wave function 
becomes for the insulating case
\begin{eqnarray}
| \: \Psi^U_I \: \rangle \: = \: \prod_{\bf i} \: \left[ \: 
\tilde F_{{\bf i}, \downarrow} \: \tilde F_{{\bf i}, \uparrow} \:
\hat F_{{\bf i}, f} \: \right] \: | \: 0 \: \rangle \: . 
\label{ins}
\end{eqnarray}
We mention that this ground-state is present only at $3 / 4$ filling, 
$p \: = \: q$, and real coupling constants in the Hamiltonian. It is 
interesting to note that for $p \: = \: q$ and real, the ${\vec k}$ 
representations for the ground-state wave functions from Eq.(\ref{ez203}) 
have no meaning since (see Eq.(\ref{ez200},\ref{ez54})) we have 
$\prod_{\vec k} \: \Delta_{\vec k} \: = \: 0$ 
for this case. This means, that the norm of the initial 
$| \: \Psi^U_g \: \rangle$ wave function from Eq.(\ref{ez17}) is not 
well defined (i.e. is zero) as expressed in the form of a product
over plaquette operators in its $\hat F^{(1)} \: \hat F^{(2)}$ part. 
However, a re-structuration in the ground-state wave function by changing 
$\hat A_{I, \sigma}$ to $\tilde F_{{\bf i}, \sigma}$ eliminates from the 
wave vector the source of the zero norm, and the ground-state from 
Eq.(\ref{ins}) becomes well defined, and its norm is finite.

From Eq.(\ref{ins}) we obtain $| \: \Psi^U_I \: \rangle \: = \: 
\prod_{\bf i} \: \hat F^{I}_{\bf i} \: | \: 0 \: \rangle$, where
\begin{eqnarray}
\hat F^{I}_{\bf i} \: = \: [ \: p^2 \: \alpha_{{\bf i}, \uparrow} \:
\hat c^{\dagger}_{{\bf i}, \downarrow} \: \hat c^{\dagger}_{{\bf i}, 
\uparrow} \: \hat f^{\dagger}_{{\bf i}, \uparrow} \: + \: p^2 \: 
\alpha_{{\bf i}, \downarrow} \: \hat c^{\dagger}_{{\bf i}, \downarrow} \:
\hat c^{\dagger}_{{\bf i}, \uparrow} \: \hat f^{\dagger}_{{\bf i},
\downarrow} \: + \: p \: \alpha_{{\bf i}, \uparrow} \: \hat f^{\dagger}_{
{\bf i}, \downarrow} \: \hat c^{\dagger}_{{\bf i}, \uparrow} \: 
\hat f^{\dagger}_{{\bf i}, \uparrow} \: + \: p \: \alpha_{{\bf i},
\downarrow} \: \hat c^{\dagger}_{{\bf i}, \downarrow} \: \hat f^{\dagger}_{
{\bf i}, \uparrow} \: \hat f^{\dagger}_{{\bf i}, \uparrow} \: ] \: .
\label{ins1}
\end{eqnarray}
Introducing the notation $| \alpha_{\bf i} |^2 \: = \: | \alpha_{{\bf i},
\uparrow} |^2 \: + \: | \alpha_{{\bf i}, \downarrow} |^2$, and 
$q_{\bf i} \: = \: [ \: ( \: 1 \: + \: p^2 \: ) \: | \alpha_{\bf i} |^2 \:
]^{-1}$, all nonzero ground-state one-particle expectation values can be 
given as
\begin{eqnarray}
&&\langle \: \hat f^{\dagger}_{{\bf i}, \sigma} \: \hat f_{{\bf i}, \sigma} \:
\rangle \: = \: q_{\bf i} \: [ \: p^2 \: | \alpha_{{\bf i}, \sigma} |^2 \:
+ \: | \alpha_{\bf i} |^2 \: ] , \quad \langle \: \hat c^{\dagger}_{{\bf i},
\sigma} \: \hat c_{{\bf i}, \sigma} \: \rangle \: = \: q_{\bf i} \: [ \:
p^2 \: | \alpha_{\bf i} |^2 \: + \: | \alpha_{{\bf i}, \sigma} |^2 \: ] \: ,
\nonumber\\
&&\langle \: \hat c^{\dagger}_{{\bf i}, \sigma} \: \hat f_{{\bf i}, \sigma} 
\: \rangle \: = \: q_{\bf i} \: p \: | \alpha_{{\bf i}, - \sigma} |^2 \: .
\label{ins2}
\end{eqnarray}
Summing up over $\sigma$ we obtain
\begin{eqnarray}
\sum_{\sigma} \langle \hat f^{\dagger}_{{\bf i},\sigma} \hat f_{{\bf i},
\sigma} \rangle \: = \: \frac{p^2 + 2}{p^2 + 1} \: , \quad
\sum_{\sigma} \langle \hat c^{\dagger}_{{\bf i},\sigma} \hat c_{{\bf i},
\sigma} \rangle \: = \: \frac{2 p^2 + 1}{p^2 +1} \: , \quad 
\sum_{\sigma} \langle \hat c^{\dagger}_{{\bf i},\sigma} \hat f_{{\bf i},
\sigma} \rangle \: = \: \frac{p}{p^2 + 1} \: .
\label{ins3}
\end{eqnarray}
Evidently, here $\langle ... \rangle \: = \: \langle \Psi^U_I | ...
| \Psi^U_I \rangle / \langle \Psi^U_I | \Psi^U_I \rangle$ holds.
Based on Eq.(\ref{ins3}) we re-obtain $N \: = \: 3 N_{\Lambda}$, and the 
ground-state expectation values of different Hamiltonian terms becomes
\begin{eqnarray}
&&\langle \: \hat T_c \: \rangle \: = \: 0 \: , \quad \langle \: \hat T_f \:
\rangle \: = \: 0 \: , \quad \langle \: \hat V \: \rangle \: = \: 0 \: , \quad
\langle \: \hat V_0 \: \rangle \: = \: \frac{2 \: p \: V_0 \: N_{\Lambda}}{1 
+ p^2} \: ,
\nonumber\\
&&\langle \: \hat E_f \: \rangle \: = \: E_f \: N_{\Lambda} \: \left( \:
1 \: + \: \frac{1}{1 + p^2} \: \right) \: , \quad
\langle \: \hat U \: \rangle \: = \: \frac{U \: N_{\Lambda}}{1 + p^2} \: .
\label{ins4}
\end{eqnarray} 
Summing up all contributions in Eq.(\ref{ins4}) we obtain for the ground-state
energy of the insulating phase
\begin{eqnarray}
\frac{E^I_0}{N_{\Lambda}} \: = \: \frac{1}{1 + p^2} \: [ \: U \: + \: E_f \:
( \: p^2 \: + \: 2 \: ) \: + \: 2 \: p \: V_0 \: ] \: .
\label{ins5}
\end{eqnarray}
Introducing $\langle \: \hat R_{loc} \: \rangle \: = \: \langle \: \hat U \:
\rangle \: + \: \langle \: \hat E_f \: \rangle \: + \: \langle \: \hat V_0 \:
\rangle$ as the contribution of the on-site (i.e. localized) Hamiltonian 
terms into the ground state energy, and $\langle \: \hat R_{mov} \: \rangle \:
= \: \langle \: \hat T_c \: \rangle \: + \: \langle \: \hat T_f \: \rangle \:
+ \: \langle \: \hat V \: \rangle$ as the contribution in the ground-state
energy of the Hamiltonian terms connected to the movement of particles within
the system, we find
\begin{eqnarray}
\langle \: \hat R_{loc} \: \rangle \: = \: E_0^I, \quad
\langle \: \hat R_{mov} \: \rangle \: = \: 0 \: ,
\label{ins5a}
\end{eqnarray}
which clearly shows that the system is completely localized.

In fact, from Eq.(\ref{ins5}) we have $E^I_0 / N_{\Lambda} \: = \: - U \: + 
\: 2 \: ( U \: + \: E_f ) \: + \: [ \: 2 \: p \: V_0 \: - \: p^2 \: ( U \:
+ \: E_f ) \: ] / ( 1 \: + \: p^2)$, which becomes 
$E^I_0 / N_{\Lambda} \: = \: - U \: + \: 2 \: ( U \: + \: E_f ) \: - \: K$ via
\begin{eqnarray}
K \: = \: p^2 \: \bar S \: , \quad V_0 \: = \: - p \: \bar S \: , \quad 
U \: + \: E_f \: = \: K \: - \: \bar S \: , \quad \bar S \: = \: 
\sum_{i = 1}^4 \: | a_{i,f} |^2 \: .
\label{ins6}
\end{eqnarray}
But, for $p \: = \: q$ and all coupling constants real, Eqs.(\ref{ins6}) can 
be directly obtained from the system of equations Eqs.(\ref{ez4},\ref{ez401a}
) . As can be seen, the ground-state energy obtained in Eq.(\ref{ins5})
gives exactly back the value deduced in Eq.(\ref{ez14}) at $N \: = \: 3 \:
N_{\Lambda}$. Taking into account from Eq.(\ref{ez42}), that $p \: = \:
V_{2,x+y} / t_{f,x+y}$, based on Eq.(\ref{ins6}), we find the surface 
$( U \: + \: E_f ) / V_0 \: = \: ( 1 \: - \: p^2 ) / p$ in the 
$\{ U, E_f, V_0 \}$ parameter space where the described solution emerges. 
The multiple conditions seen in Eq.(\ref{ez42}) relating the hopping matrix
elements and neighboring hybridizations to the $p$ parameter can be achieved
in the simplest way in the case of the non-distorted unit cell. As can be 
seen, $\delta \: = \: 0$ for $p \: = \: q$ and $p \: = p^{*}$, represents a
possible physical solution for the analyzed problem.

Before continuing, some observations has to be made here. 
First of all, from Eq.(\ref{ins3}) one can see that the sum over $\sigma$
of the nonzero ground-state expectation values presented in Eq.(\ref{ins2})
are independent on the $\alpha_{{\bf i}, \sigma}$ coefficients introduced by
$\hat F^{(3)}$ in Eq.(\ref{ez16}). This implies that the results given in
Eq.(\ref{ins4}) are valid also for the most general form of the ground-state
wave function expressed with $\hat F^{(3)}$ constructed via Eq.(\ref{ez16a})
instead of Eq.(\ref{ez16}). We underline that the independence on the
$\alpha_{{\bf i}, \sigma}$ coefficients of the ground-state expectation 
values summed up over $\sigma$ is a general property of the ground-state wave 
function, and is valid not only in the insulating case described here
(see Appendix).

Secondly, we mention that doping the system, the properties described here
will be destroyed. The reason for this is that introducing electrons above 
$3 / 4$ filling, the number of particles per site will not be constant 
along the whole lattice, and as a consequence, the ground-state expectation 
values of the kinetic energy terms become to be nonzero.  

\subsection{The non-symmetric case}
 
Both $p$ and $q$ being considered pure real, we concentrate now on the 
possibility of a $p \: \ne \: q$ solution. We arrive to this inequality for 
example taking into account distorted unit cell, which gives usually the 
$p \: \ne \: q$ condition. The $\delta$ value introduced after 
Eq.(\ref{ez46}) is nonzero, so the state we are analyzing is clearly a 
non-localized state. Solving however for this case Eq.(\ref{ez4}), and 
calculating the norm of the ground state wave function $| \: \Psi^U_g \:
\rangle$ from Eq.(\ref{ez17}), we obtain zero value. 

Indeed, solving the system of equations Eqs.(\ref{ez4},\ref{ez401}) for 
$p \: \ne \: q$ and both real (see Eq.(\ref{ez40c})), we finally obtain
\begin{eqnarray}
&&| \: X_c \: |^2 \: = \: K \: - \epsilon_{\vec k}^c \: = \: 
4 \: p^2 \: | t_{f,x+y} | \: ( \: {\cal{C}}_{1, s, {\vec k}} \: + \:
\frac{s \: q \: r}{p} \: {\cal{C}}_{2, s, {\vec k}} \: )^2 \: ,
\nonumber\\
&&| \: X_f \: |^2 \: = \: K \: - \: \epsilon_{\vec k}^f \: = \:
4 \: | t_{f,x+y} | \: ( \: {\cal{C}}_{1, s, {\vec k}} \: + \: r \: s \: 
{\cal{C}}_{2, s, {\vec k}} \: )^2 \: ,
\nonumber\\
&&V_{\vec k} \: = \: - \: 4 \: p \: | t_{f,x+y} | \: ( \: {\cal{C}}_{1, s,
{\vec k}} \: + \: \frac{s \: q \: r}{p} \: {\cal{C}}_{2, s, {\vec k}} \: ) 
\cdot ( \: {\cal{C}}_{1, s, {\vec k}} \: + \: r \: s \: {\cal{C}}_{2, s,
{\vec k}} \: ) \: , 
\nonumber\\
&&K \: = \: 2 \: p^2 \: | t_{f,x+y} | \: \left( \: 1 \: + \: \frac{q^2 \:
r^2}{p^2} \: \right), \quad V_0 \: = \: - 2 \: p \: | t_{f,x+y} | \: \left( \:
1 \: + \: r^2 \: \frac{q}{p} \: \right) \: ,
\label{ez54}
\end{eqnarray}
where $s \: = \: t_{f,x} / t_{f,y}$, $r \: = \: t_{f,x} / ( \: 2 \: t_{f,x+y}
\: )$, and, for $s \: > \: 0$ we have ${\cal{C}}_{i, s, {\vec k}} \: = \:
\cos[ \: \frac{\vec k}{2} \cdot ( \: {\vec x} \: + \: (-1)^{i+1} \: {\vec y} 
\: ) \: ]$ and for $s \: < \: 0$, ${\cal{C}}_{i, s, {\vec k}} \: = \: \sin[ \:
\frac{\vec k}{2} \cdot ( \: {\vec x} \: + \: (-1)^{i+1} \: {\vec y} \: ) \: 
]$, respectively. We mention that for the case of real coupling constants in 
the Hamiltonian and $p \: = \: p_1 \: = \: p_2 \: \ne \: q \: = \: q_1 \:
= \: q_2$ presented here, the system of equations Eqs.(\ref{ez4},\ref{ez401}) 
admits solutions only for $- \: s \: t_{f,x+y} \: > \: 0$, $sign[ \: t_{f,x+y}
\: ] \: = \: sign[ \: t_{f,y-x} \: ]$ and $t_{f,x}^2 \: = \: t_{f,y}^2$. From
Eq.(\ref{ez54}) it can be seen that there exist at least one ${\vec k}$ value,
for which $\Delta_{\vec k} \: =  \: 0$ (for example, $k_x \: = \: k_y \: = \:
0$ for $s \: < \: 0$, or $k_y \: = \: 0, \: k_x \: = \: \pi$ for $s \: > \:
0$ ). So indeed, the norm of the ground-state wave function is vanishing, it 
is not possible to remove this property as in $p \: = \: q$ case, and as a 
consequence, the solution is not proper for describing the presented 
situation.

As a conclusion, we can see that for pure real coupling constants in the 
Hamiltonian, the described solution at $3 / 4$ filling represents a completely
localized state. The part of the wave function $| \: \Psi^U_g \: \rangle$ 
that gives nonzero norm introduces on every site of the lattice rigorously 
the same number of electrons. Given by this reason, a hopping 
$\hat c^{\dagger}_{{\bf i}, \sigma} \: \hat c_{{\bf j}, \sigma}$,
$\hat f^{\dagger}_{{\bf i}, \sigma} \: \hat f_{{\bf j}, \sigma}$, or
$\hat c^{\dagger}_{{\bf i}, \sigma} \: \hat f_{{\bf j}, \sigma}$, with 
${\bf i} \: \ne \: {\bf j}$ creates a state orthogonal to 
$| \: \Psi^U_g \: \rangle$. As a consequence, all ground-state expectation 
values connected to the movement of particles within the system (i.e. kinetic
energy terms and neighboring hybridizations) are zero. In this respect, the 
obtained state represents a paramagnetic Mott insulator, and the ground-state
energy can be obtained as a sum of the ground-state expectation values of the 
on-site terms from the Hamiltonian. 

Before continuing, we would like to underline the extreme sensitivity of the 
solutions contained in Eqs.(\ref{ez4},\ref{ez401a}) to lattice distortions. 
In specially we stress, that in the case of the undistorted unit cell (i.e. 
$t_{b,x} \: = \: t_{b,y}, \: t_{b,x+y} \: = \: t_{b,y-x}, \:
V_{1,x} \: = \: V_{1,y}, \: V_{2,x+y} \: = \: V_{2,y-x}$) the system of 
equations Eqs.(\ref{ez4}, \ref{ez401a}) admits only pure real solutions for 
$p$ and $q$, i.e. a completely localized ground-state.

\section{The non-localized solution}

In order to obtain another type of solution than that presented in the 
previous Section, we must consider the hybridization coupling constants 
imaginary, the Hamiltonian remaining hermitian. We further consider in this 
Section all hopping matrix elements real and $V^{cf}_{\gamma} \: = \: 
V^{fc}_{\gamma}$ for all hybridization matrix elements $\gamma$.
Again, as in the case of the completely localized solution, two cases emerge,
namely $|p| \: \ne \: |q|$, and $|p| \: = \: |q|$, which will be analyzed 
separately.  

\subsection{The possible solutions for imaginary p and q}

We study now the solutions allowed by Eqs.(\ref{ez4},\ref{ez401a}) in case of
imaginary $p$ and $q$. The first group of solutions obtained, emerge at 
$|p| \: \ne \: |q|$, case that will be denoted as non-symmetric below.

\subsubsection{The non-symmetric case}

Introducing the notations from Eq.(\ref{ez42}) and being interested
only in situations described by Eq.(\ref{ez46}), for $|p| \: \ne \: |q|$ the 
solutions allowed by Eqs.(\ref{ez4},\ref{ez401a}) are characterized by five
independent and free starting parameters, namely
$V_{2,x+y} \: = \: i \: \bar V_{2,x+y}, \: V_{2,y-x} \: = \: i \: 
\bar V_{2,y-x}, \: t_{f,x+y}, \: t_{f,y-x}$, and $U$, where $\bar V_{\gamma}$
are pure real variables. Solutions are obtained for  $| t_{f,x} | \: = \: |
t_{f,y} | \: = \: 2 \: \sqrt{t_{f,x+y} \: t_{f,y-x}}$, $\theta \: = \: - \:
t_{f,x+y} \: t_{f,x} / t_{f,y} \: > \: 0$ and gives $V_{1,x} \: = \: ( \: p
\: + \: q^{*} \: ) \: t_{f,x} / 2$, $V_{1,y} \: = \: ( \: p \: + \: q \: ) \:
t_{f,y} / 2$, $V_0 \: = \: 0$. We obtain $a_{1,f} \: = \: \chi e^{i\phi}, \:
a_{2,f} \: = \: 2 \: \chi \: ( \: t_{f,y-x} / t_{f,y} \: ) \: e^{i \phi}, \:
a_{3,f} \: = \: \chi \: ( \: t_{f,x} / t_{f,y} \: ) \: e^{i\phi}, \:
a_{4,f} \: = \: 2 \: \chi \: ( \: t_{f,y-x} / t_{f,x} \: ) \: e^{i \phi}$,
where $\chi \: = \: \sqrt{\theta}$, and $\phi$ is an arbitrary phase. From 
this, $K \: = \: 2 \: | \chi |^2 \: ( \: |p|^2 \: + \: |q|^2 \: | 2 \: 
t_{f,y-x} / t_{f,y} |^2 \: ), \: \: K_f \: = \: K \: - \: \tilde E_f \: = \: 
2 \: |\chi|^2 \: ( \: 1 \: + \: | 2 \: t_{f,y-x} / t_{f,y} |^2 \: )$. 
Introducing the notations $r \: = \: t_{f,x} / ( \: 2 \: t_{f,x+y} )$, $s \: =
\: t_{f,y} / t_{f,x}$, for $| X_c( {\vec k} ) |^2$ we obtain
\begin{eqnarray}
| X_c( {\vec k} ) |^2 \: = \: 4 \: |p|^2 \: |t_{f,x+y}| \: \left( \: 
{\cal C}_{1, -s, {\vec k}} \: + \: \frac{s \: q \: r}{p} \: {\cal C}_{2,
-s, {\vec k}} \: \right) \: ,
\label{ez150}
\end{eqnarray}
where ${\cal C}_{i, s, {\vec k}}$ has been introduced in Eq.(\ref{ez54}). 
Since for a well defined non-zero norm we have to have $| X_c({\vec k}) |^2 
\: > \: 0$ for all ${\vec k}$ (see for example Eq.(\ref{ez211a}), or the 
explication presented after Eq.(\ref{ez203})), Eq.(\ref{ez150}) shows that
the non-symmetric case described here fails to represent a proper physical
solution.

\subsubsection{The symmetric case}

From mathematical point of view the symmetric  $|p| \: = \: |q|$ solution is 
more complicated than the non-symmetric one. For this case two situations 
emerge, namely $p \: = \: q$, and $p \: = \: q^{*}$, respectively. These two 
situations are however physically equivalent, and can be obtained each from 
other by a rotation with $\pi/2$ of the system of coordinates. Because of 
this reason, we have to analyze in detail only one of them, namely the 
$p \: = \: q$, $p \: = \: - p^{*}$. As for the non-symmetric solutions, we 
have $V_{\gamma} \: = \: i \: \bar V_{\gamma}$ for all hybridization matrix 
elements.

The $p \: = \: q$ solution emerge only for distorted unit cell. Five 
parameters can be independently chosen, namely $t_{f,x+y}, \: t_{f,y-x}, \:
V_{2,x+y}, \: V_0,$  and $U$, so that $t_{f,x+y} \: \ne \: t_{f,y-x}$ and 
$sign ( t_{f,x+y} ) \: = \: sign ( t_{f,y-x } )$. The solution gives via the
$p \: = \: V_{2,x+y}/t_{f,x+y}$ parameter the relations $t_{c,x+y} \: = \:
p^2 \: t_{f,x+y}$, $t_{c,y-x} \: = \: p^2 \: t_{f,y-x}$, $t_{c,x} \: = \:
|p|^2 \: t_{f,x}$, $t_{c,y} \: = \: - |p|^2 \: t_{f,y}$, $V_{1,x} \: = \: 0$,
$V_{1,y} \: = \: p \: t_{f,y}$, $V_{2,y-x} \: = \: p \: t_{f,y-x}$, 
$t^2_{f,x} \: = \: 4 \: t_{f,x+y} \: t_{f,y-x}$, and $t_{f,y} \: = \: - 
t_{f,x} \: [ \: sign( t_{f,x+y} ) \: ] \: [ \: 1 \: + \: ( \bar \theta)^2 \:
]^{1/2}$ so that $ - \: t_{f,x} \: t_{f,y} / t_{f,x+y} \: > \: 0$, $ - r \:
t_{f,y} \: > \: 0$, where $r \: = \: t_{f,x} / ( \: 2 \: t_{f,x+y} \: )$, 
$\bar \theta \: = \: \bar V_0 \: t_{f,x+y} / [ \: 2 \: \bar V_{2,x+y} \: 
( \: t_{f,y-x} \: - \: t_{f,x+y} \: ) \: ]$. For the $a_{i,f}$ coefficients 
we find $| a_{1,f} |^2 \: = \: w \: - \: u, \: | a_{2,f} |^2 \: = \: - \: 
r^2 \: ( \: w  \: - \: u \: ), \: | a_{3,f} |^2 \: = \: - \: w \: - \: u, \:
| a_{4,f} |^2 \: = \: r^2 \: ( \: w \: - \: u \: ), $ where $w \: = \: V_0
/ [ \: 2 \: p \: ( \: r^2 \: - \: 1 \: ) \: ]$ and $u \: = \: t_{f,y} / ( \: 
2 \: r \: )$. We have to consider $|p|^2 \: > \: 1$ and $r^2 \: \ne \: 1$, 
the solution being present in the $\{ U, E_f, V_0\}$ parameter space on the 
surface\cite{P0}
\begin{eqnarray}
U \: + \: E_f \: = \: \frac{| t_{f,y} |}{| r |} \: ( \: |p|^2 \: - \: 1 \: )
\: ( \: r^2 \: + \: 1 \: ) \: .
\label{ez151}
\end{eqnarray}
Introducing the notation $s \: = \: - t_{f,x+y} / | t_{f,x+y} |$, for $z \:
= \: t_{f,x} / t_{f,y}$ we obtain $z \: = \: s / \sqrt{1 \: + \: 
(\bar \theta)^2} $, so $0 \: < \: |z| \: < \: 1$ emerge. For 
$| X_c( {\vec k} ) |^2$, using the notation
${\cal D}_{\vec k} \: = \: ( \: {\cal C}_{1, -s, {\vec k}} \: + \: r \: s \: 
{\cal C}_{2, -s, {\vec k}} \: )^2$,  we find 
\begin{eqnarray}
| X_c({\vec k}) |^2 \: = \: 4 \: |p|^2 \: |t_{f,x+y}| \: \left[ \: 
\: {\cal D}_{\vec k} \: + \: \frac{1 - |z|}{ 2 |z|} \: [ \: 1 \: + \: r^2 \: 
- \: 2 \: r \: \cos(k_y) \: ] \: \right] \: .
\label{ez152}
\end{eqnarray}
For $|r| \: \ne \: 1$ the expression from Eq.(\ref{ez152}) is strictly 
positive, so the norm of the ground-state wave function is nonzero and well 
defined. In this case we obtain besides $| X_c({\vec k}) |^2 \: > \: 0$ as 
well $\Delta_{\vec k} \: > \: 0$, since $| X_c({\vec k}) |^2$ and $| X_f(
{\vec k}) |^2$ change their sign at the same ${\vec k}$ value. As explained
in Sec. III., in the present case $\Delta > 0$, which physically means that 
the diagonalized bands from Eq.(\ref{ez40}) are non-intersecting and 
completely separated.

From mathematical point of view, at the level of $| X_c({\vec k}) |^2$, 
the main difference between the non-symmetric and symmetric case presented 
in this Section arises from the fact that for $p \: \ne \: q$ we have 
$| a_{1,c} | \: = \: | a_{3,c} |, \: | a_{2,c} | \: = \: | a_{4,c} |$, and 
the phase of $a_{1,c}, \: a_{3,c}$, (or $a_{2,c}, \: a_{4,c}$), differs only 
by $0$ or $\pi$. Such properties are not present for $p \: = \: q$.
In view of Eq.(\ref{ez40b}), this means that in the non-symmetric case, 
coupling together $(a'_{1,c}, \: a'_{3,c})$, and $(a'_{2,c}, \: a'_{4,c})$ 
in the expression of $|X_c({\vec k})|^2 \: = \: K \: - \: \epsilon^c_{
\vec k}$, the trigonometric factors ${\cal C}_{i, s, {\vec k}}$ automatically
arise, leading to Eq.(\ref{ez150}). For the symmetric case, a such type of 
direct term grouping in $| X_c({\vec k}) |^2$ is no more possible.

\subsection{Ground-state expectation values}

Once in conditions presented in Sec.VI.A.2. the norm of the ground-state wave
function is well defined and $|X_c({\vec k})|^2 \: > \: 0, \: \prod_{\vec k} 
\: \Delta_{\vec k} \: > \: 0$, we have find a completely different solution 
in comparison with the ground-state presented in Sec. V. Being motivated by 
the interest to find the physical properties of the system in the analyzed 
case, we can start now the calculation of all ground-state expectation values
of interest. Since the study can now be easier done in the ${\vec k}$ 
coordinates, we are presenting the calculations using the ${\vec k}$ 
representation of the ground-state wave function and Hamiltonian described in
Sec.III.B.

The deduction of the main expectation values based on which the physical
interpretations are made can be followed using the presented Appendix in
their full generality in the case of arbitrary $\alpha_{{\bf i}, \sigma}$
coefficients. However, in order to be easier for the reader to follow the
main steps of the deduction, we are presenting below a simplified version of 
the calculation obtained in the case of site independent $\alpha_{{\bf i},
\sigma}$ coefficients, which leads to the same results. In fact, the Appendix
shows for example, that rigorously, in order to calculate ground-state 
expectation values of different Hamiltonian terms (i.e. expectation values 
summed over the spin index), is enough to consider site independence for the 
$\alpha_{{\bf i}, \sigma}$ coefficients entering in the $\hat F^{(3)}$ 
operator present in the ground-state wave function. We mention, that from 
physical point of view, the difference between $\alpha_{{\bf i}, \sigma} \: =
\: \alpha_{\sigma}$, and site dependent $\alpha_{{\bf i}, \sigma}$, is that 
the first case describes the maximal total spin $S$ part of the ground-state
wave function only, and the second case takes into account the full expression
of the ground-state wave function. 

Introducing for the $\alpha_{{\bf i}, \sigma} \: = \: \alpha_{\sigma}$ case 
the notation $|\alpha|^2 \: = \: \sum_{\sigma} \: | \: \alpha_{\sigma} \: 
|^2$, the $\hat F^{(3)}$ operator becomes $\hat F^{(3)} \: = \: \prod_{\bf i} 
\: \hat F_{\bf i}^{(3) \dagger}$ where
\begin{eqnarray}
{\hat F}_{\bf i}^{(3) \dagger} \: = \: |\alpha|^{-1} \: (
\: \alpha_{\uparrow} \: {\hat f}^{\dagger}_{{\bf i}, \uparrow} \: + \:
\alpha_{\downarrow} \: {\hat f}^{\dagger}_{{\bf i}, \downarrow} ) \: ,
\label{ez210}
\end{eqnarray}
the coefficients $\alpha_{\sigma}$ being arbitrary. In the ${\vec k}$
space the contribution of $\hat F^{(3)}$ (taking into account normalized
factors) becomes in this case $\hat F^{(3)} \: = \: \sum_{\vec k} \:
\hat F^{(3)}_{\vec k}$, where
\begin{eqnarray}
\hat F^{(3)}_{\vec k} \: = \: \sum_{\sigma} \: \frac{\alpha_{\sigma}}{
|\alpha|} \: \hat f^{\dagger}_{{\vec k}, \sigma} \: ,
\label{ez211}
\end{eqnarray}
and for the norm of the ground-state wave function from Eq.(\ref{ez203})
we obtain
\begin{eqnarray}
\langle \: \Psi^U_g \: | \: \Psi^U_g \: \rangle \: = \: \prod_{\vec k} \:
| X_c({\vec k}) |^2 \: .
\label{ez211a}
\end{eqnarray}
In the case of $3 / 4$ filling, the desired expectation values becomes
(we have here $\langle ... \rangle = \langle \Psi^U_g | ... | \Psi^U_g \rangle
/ \langle \Psi^U_g | \Psi^U_g \rangle $)
\begin{eqnarray}
&&\langle \: \hat c^{\dagger}_{{\vec k}, \sigma} \: \hat c_{{\vec k},\sigma}
\: \rangle = \frac{|\alpha|^2 |X_c({\vec k})|^2 \: + \: |\alpha_{\sigma}|^2 
|X_f({\vec k})|^2}{|\alpha|^2 \Delta_{\vec k}} \: , \quad
\langle \: \hat f^{\dagger}_{{\vec k},\sigma} \: \hat f_{{\vec k},\sigma}
\: \rangle = \frac{|\alpha|^2 |X_f({\vec k})|^2 \: + \: |\alpha_{\sigma}|^2 
|X_c({\vec k})|^2}{|\alpha|^2 \Delta_{\vec k}} \: , 
\nonumber\\
&&\langle \: \hat f^{\dagger}_{{\vec k},\sigma} \: \hat c_{{\vec k},\sigma}
\: \rangle \: = \: - \: \frac{ V_{\vec k} \: |\alpha_{-\sigma}|^2 }{
|\alpha|^2 \: \Delta_{\vec k}} \: .
\label{ez212}
\end{eqnarray}
Effectuating the sum over the spin index in Eq.(\ref{ez212}), as expected
from the Appendix, the numerical coefficients $\alpha_{\sigma}$ disappear 
from the expectation values
\begin{eqnarray}
&&\sum_{\sigma} \: \langle \hat c^{\dagger}_{{\vec k},\sigma} 
\hat c_{{\vec k},\sigma} \rangle = \frac{2 \: |X_c({\vec k})|^2 + 
|X_f({\vec k})|^2}{\Delta_{\vec k}} \: ,\quad
\sum_{\sigma} \: \langle \hat f^{\dagger}_{{\vec k},\sigma} 
\hat f_{{\vec k},\sigma} \rangle = \frac{2 \: |X_f({\vec k})|^2 +  
|X_c({\vec k})|^2}{\Delta_{\vec k}} \: , 
\nonumber\\
&&\sum_{\sigma} \: \langle \: \hat f^{\dagger}_{{\vec k},\sigma} \: 
\hat c_{{\vec k},\sigma} \: \rangle \: = \: - \: \frac{ V_{\vec k} }{ 
\Delta_{\vec k}} \: .
\label{ez213}
\end{eqnarray}
Comparing with Eqs.(\ref{a17},\ref{a18}), we see that the expectation values
from Eq.(\ref{ez213}) are true for arbitrary $\alpha_{{\bf i},\sigma}$, so
are correct also in the case of using the general Eq.(\ref{ez16a}) instead of 
Eq.(\ref{ez16}) into $\hat F^{(3)}$. The ground-state expectation values of
different Hamiltonian terms become
\begin{eqnarray}
&&\langle \hat T_c \rangle = \sum_{\vec k} \frac{\epsilon_{\vec k}^c}{
\Delta_{\vec k}} ( 2 \: |X_c({\vec k})|^2 + |X_f({\vec k})|^2 ) \: , \quad
\langle \hat T_f \rangle = \sum_{\vec k} \frac{(\epsilon_{\vec k}^f - 
\tilde E_f)}{
\Delta_{\vec k}} ( 2 \: |X_f({\vec k})|^2 + |X_c({\vec k})|^2 ) \: ,
\nonumber\\
&&\langle \hat V \rangle = - \sum_{\vec k} \frac{2 \: |V_{\vec k}|^2 - \:
( V_{\vec k}^{*} V_0 + V^{*}_0 V_{\vec k} ) }{\Delta_{\vec k}} \: , \quad
\langle \hat U \rangle \: = \: U \: \sum_{\vec k} \: \frac{|X_f({\vec k})|^2 }
{\Delta_{\vec k}} \: ,
\nonumber\\
&&\langle \hat V_0 \rangle \: = \: - \sum_{\vec k} \frac{V_{\vec k}^{*} V_0 +
V^{*}_0 V_{\vec k}}{\Delta_{\vec k}} \: , \quad
\langle \hat E_f \rangle \: = \: E_f \: \sum_{\vec k} \: 
\frac{2 \: |X_f({\vec k})|^2 + |X_c({\vec k})|^2}{\Delta_{\vec k}} \: .
\label{ez214}
\end{eqnarray} 
To have more insight about the physical behavior of the system, all
ground-state expectation values at $U \: > \: 0$ relevant for our study can be
explicitly expressed from Eq.(\ref{ez214}). The analysis of the described 
state first of all shows that $N \: = \: \sum_{{\vec k}, \sigma} \: [ \:
\langle \: ( \: \hat f^{\dagger}_{{\vec k}, \sigma} \: \hat f_{{\vec k},
\sigma} \: + \: \hat c^{\dagger}_{{\vec k}, \sigma} \: \hat c_{{\vec k},
\sigma} \: ) \: \rangle \: ] \: = \: 3 \: N_{\Lambda}$, as must be for 
$3 / 4 $ filling from Eq.(\ref{ez213}). Secondly, summing up all 
contributions from Eq.(\ref{ez214}), we obtain for the ground-state energy 
$E_g \: = \: \sum_{\vec k} \: ( \: \Delta_{\vec k}^{-1} \: ) \: [ \:
\epsilon_{\vec k}^c \: ( \: 2 \: | X_c({\vec k}) |^2 \: + \: | X_f({\vec k})
|^2 \: ) \: + \: \epsilon_{\vec k}^f \: ( \: 2 \: | X_f({\vec k}) |^2 \:
+ \: | X_c({\vec k}) |^2 \: ) \: - \: 2 \: | V_{\vec k} |^2 \: - \: U \:
( \: | X_c({\vec k}) |^2 \: + \: | X_f({\vec k}) |^2 \: ) \: ]$. Taking into
account from Eq.(\ref{ez52}) $\Delta_{\vec k} \: = \: ( \: | X_c({\vec k})
|^2 \: + \: | X_f({\vec k}) |^2 \: )$, this expression becomes $E_g \: = \: 
2 \: \sum_{\vec k} \: ( \: \epsilon^c_{\vec  k} \: + \: \epsilon^f_{\vec k} \:
) \: - \: \sum_{\vec k} \: ( \: \Delta_{\vec k}^{-1} \: ) \: [ \: 
\epsilon_{\vec k}^c \: | X_f({\vec k}) |^2 \: + \: \epsilon_{\vec k}^f \:
| X_c({\vec k}) |^2 \: - \: 2 \: | V_{\vec k} |^2 \: ] \: - \: U \:
N_{\Lambda}$. The first sum in $E_g$ gives  $2 \: \tilde E_f \: N_{\Lambda}$ 
(we note at this step that at $U \: \ne \: 0$ we have $\tilde E_f$ instead of
$E_f$ in the relation $\epsilon^f_{{\vec k},\sigma} \: = \: E_f \: + \:
\varepsilon^f_{{\vec k},\sigma}$ following Eq.(\ref{ez26})). Using 
Eqs.(\ref{ez38},\ref{ez52},\ref{ez53}), the second sum becomes $- \: K \:
N_{\Lambda}$. As a consequence, the ground-state energy obtained from 
Eq.(\ref{ez214}) is $E_g / N_{\Lambda} \: = \: 2 \: E_f \: + \: U \: - \: K$,
which is exactly the value $E^U_0 / N_{\Lambda}$ given by Eq.(\ref{ez14}) 
for $N \: = \: 3 \: N_{\Lambda}$. 

Deducing now from Eq.(\ref{ez213}) the total momentum distribution function
$n_{\vec k} \: = \: \sum_{b, \sigma} \: \langle \: \hat n^b_{{\vec k}, 
\sigma} \: \rangle$, where $b = c,f$ , we obtain $n_{\vec k} \: = \: 3$, i.e.
a completely uniform and continuous behavior for the whole ${\vec k}$-space. 
This value of $n_{\vec k}$ contains however also the contribution of the 
lower band, which being completely filled up, is $2$ for every $({\vec k}, \: 
\sigma)$ state. This can be seen as well from  Eq.(\ref{a16}) of the Appendix,
taking into consideration that $\hat f^{\dagger}_{{\vec k}, \sigma} \:
\hat f_{{\vec k}, \sigma} \: + \: \hat c^{\dagger}_{{\vec k}, \sigma} \:
\hat c_{{\vec k}, \sigma} \: = \: \hat B^{\dagger}_{{\vec k}, 1, \sigma} \:
\hat B_{{\vec k}, 1, \sigma} \: + \: \hat B^{\dagger}_{{\vec k}, 2, \sigma} 
\: \hat B_{{\vec k}, 2, \sigma}$, writing the general ground state as 
$| \: \Psi^U_g \: \rangle \: = \: \sum_{\bfsigma} \: a_{\bfsigma} \:
| \: \Psi^U_{g,\bfsigma} \: \rangle$, where $a_{\bfsigma}$ are arbitrary 
coefficients, taking into consideration that as shown in Eq.(\ref{ez208}),
$\hat B^{\dagger}_{{\vec k}, 2, \sigma}$ creates a particle in the lower 
band, and obtaining $\langle \: \sum_{\sigma} \: \hat B^{\dagger}_{{\vec k},
2, \sigma} \: \hat B_{{\vec k}, 2, \sigma} \: \rangle \: = \: 2$ independent 
on the $({\vec k}, \: \sigma)$ indices. As a consequence, $n_{\vec k} \: = 
\: 3$ means in fact, that for the upper band (denoted by $1$) we have 
$n_{\vec k}^{(1)} \: = \: 1$ independent on ${\vec k}$. This information
is also contained in Eq.(\ref{a15}) of the Appendix, which gives 
$\langle \: \sum_{\sigma} \: \hat B^{\dagger}_{{\vec k}, 1, \sigma} \: 
\hat B_{{\vec k}, 1, \sigma} \: \rangle \: = \: 1$, where as shown through 
Eq.(\ref{ez208a}), $\hat B^{\dagger}_{{\vec k}, 1, \sigma}$ creates a 
particle in the upper band. Using Eqs.(\ref{a16}, \ref{a20}) we have even 
$\langle \: \hat B^{\dagger}_{{\vec k}, 1, \sigma} \: \hat B_{{\vec k}, 1,
\sigma} \: \rangle \: = \: 1 / 2$, $\langle \: \hat B^{\dagger}_{
{\vec k}, 2, \sigma} \: \hat B_{{\vec k}, 2, \sigma} \: \rangle \: = \: 1$, 
which via Eq.(\ref{ez208b}) can be transformed into the $n_{\vec k}^c$ and 
$n_{\vec k}^f$ momentum distribution functions defined for the starting
operators $\hat f$ and $\hat c$.

We note at this step, that even the simplified calculation presented in 
Eq.(\ref{ez212}) gives back all essential features presented above since
it leads to $\langle \: \hat B^{\dagger}_{{\vec k}, 2, \sigma} \: 
\: \hat B_{{\vec k}, 2, \sigma} \: \rangle \: = \: 1$, $\langle \: \sum_{
\sigma} \: \hat B^{\dagger}_{{\vec k}, 1, \sigma} \: \hat B_{{\vec k}, 1,
\sigma} \: \rangle \: = \: 1$, $\langle \: \hat B^{\dagger}_{{\vec k}, 1,
\sigma} \: \hat B_{{\vec k}, 1, \sigma} \: \rangle \: = \: const$. The same 
results are re-obtained in case of ${\vec k}$ dependent $\alpha_{{\vec k},
\sigma}$ coefficients in Eq.(\ref{ez211}). 

Since $\Delta_{\vec k} \: > \: 0$ in the studied case, $n_{\vec k}$
and all individual 
contributions in $n_{\vec k}$ listed above are continuous together with 
their derivatives of any order in the whole momentum space. As a consequence,
the system is in a non-Fermi liquid (NFL) state. From physical point of view,
this property is clearly given by the presence at $U \: > \: 0$ of a 
partially filled completely flat upper band, which is not hybridized with 
the lower band that contains dispersion.
 
In order to further analyze the described state three integrals must be
introduced
\begin{eqnarray}
I_0 \: = \: \sum_{\vec k} \: \frac{| X_f({\vec k}) |^2}{\Delta_{\vec k}} \:, 
\quad I_1 \: = \: \sum_{\vec k} \: \frac{| X_f({\vec k}) |^2  \: |
X_c({\vec k}) |^2}{ \Delta_{\vec k} } \: , \quad
I_2 \: = \: \sum_{\vec k} \: \frac{V_{\vec k}}{\Delta_{\vec k}} \: .
\label{ez215}
\end{eqnarray}
Using Eq.(\ref{ez215}), from Eq.(\ref{ez214}) we now find
\begin{eqnarray}
&&\langle \: \hat T_c \: \rangle \: = \: - \: K \: I_0 \: + \: I_1 , \quad
\langle \: \hat T_f \: \rangle \: = \: - \: ( \: K \: - \: E_f \: - \: U \:
) \: N_{\Lambda} \: + \: ( \: K \: - \: E_f \: - \: U \: ) \: I_0 \: + \:
I_1 \: ,
\nonumber\\
&&\langle \: \hat V \: \rangle \: = \: - \: 2 \: I_1 \: + \: ( \: V_0^{*} \:
I_2 \: + \: V_0 \: I_2^{*} \: ) , \quad \langle \: \hat U \: \rangle \:
= \: U \: I_0 \: ,
\nonumber\\
&&\langle \: \hat V_0 \: \rangle \: = \: - \: ( \: V_0^{*} \: I_2 \: + \: 
V_0 \: I_2^{*} \: ) , \quad \langle \: \hat E_f \: \rangle \: = \: E_f \:
N_{\Lambda} \: + \: E_f \: I_0 \: ,
\label{ez216}
\end{eqnarray} 
where, starting from Eq.(\ref{ez215}) $I_0 \: > \: 0$, $I_1 \: > \: 0$, and 
based on Eqs.(\ref{ez4}, \ref{ez401a}) we have $K \: > \: 0$, $( \: K \: - \:
E_f \: - \: U \: ) \: > \: 0$, respectively. Introducing as in 
Eq.(\ref{ins5a}) the contribution of the on-site, and hopping-type Hamiltonian
terms into the ground-state energy, we get $\langle \: \hat R_{loc}
\: \rangle \: = \: E_f \: N_{\Lambda} \: + \: \tilde E_f \: I_0  \: - \:
( \: V_0^{*} \: I_2 \: + \: V_0 \: I_2^{*} \: )$, and $\langle \: \hat R_{mov}
\: \rangle \: = \: ( \: \tilde E_f \: - \: K \: ) \: N_{\Lambda} \: - \:
\tilde E_f \: I_0 \: + \: ( \: V_0^{*} \: I_2 \: + \: V_0 \: I_2^{*} \: )$ 
from where, as expected, $E^U_0 \: = \: \langle \: \hat R_{loc} \: \rangle \:
+ \: \langle \: \hat R_{mov} \: \rangle \: = \: ( \: 2 \: E_f \: + \: U \: - 
\: K \: ) \: N_{\Lambda}$ is re-obtained. From here one can write
\begin{eqnarray}
&&\langle \: \hat R_{loc} \: \rangle \: = \: E^U_0 \: + \: {\cal J}  \: , 
\quad \langle \: \hat R_{mov} \: \rangle \: = \: - \: {\cal J} \: , 
\nonumber\\
&& {\cal J} \: = \: ( \: K \: - \: E_f \: - \: U \: ) \: N_{\Lambda} \: + \:
( \: E_f \: + \: U \: ) \: I_0 \: - \: ( \: V_0^{*} \: I_2 \: + \: V_0 \:
I_2^{*}) \: .
\label{ez217}
\end{eqnarray}
The study of ${\cal J}$ shows that ${\cal J} \: > \: 0$ holds. In order to 
see this, via $\Delta_{\vec k} \: = \: |X_c({\vec k})|^2 \: + \: 
|X_f({\vec k})|^2$, we obtain ${\cal J} \: = \: \sum_{\vec k} \: 
P_{\vec k} / \Delta_{\vec k}$, where $P_{\vec k} \: = \: K \: ( \: K \: - \:
\epsilon^f_{\vec k} \: ) \: + \: ( \: K \: - \: \tilde E_f \: ) \: ( \: K \:
- \: \epsilon^c_{\vec k} \: ) \: - \: ( \: V_0^{*} \: V_{\vec k} \: + \:
c.c \: )$. Using now Eq.(\ref{ez52}), we get $P_{\vec k} \: = \: ( \: K \:
- \: \tilde E_f \: ) \: |X_c({\vec k})|^2 \: + \: K \: |X_f({\vec k})|^2 \:
+ \: [ \: X_c({\vec k}) \: X^{*}_f({\vec k}) \: V^{*}_0 \: + \: c.c \: ]$. 
Based on this relation, introducing the notations $d_{i,c} \: = \: a_{i,c} \:
X^{*}_f({\vec k}), \: d_{i,f} \: = \: a_{i,f} \: X^{*}_c({\vec k})$, and 
using Eqs.(\ref{ez4},\ref{ez401a}) we find
\begin{eqnarray}
P_{\vec k} \: = \: \sum_{i = 1}^4 \: \left[ \: | d_{i,f} |^2 \: + \:
| d_{i,c} |^2 \: - \: d_{i,c} \: d^{*}_{i,f} \: - \: d_{i,f} \: d^{*}_{i,c} 
\: \right] \: = \: \sum_{i = 1}^4 \: | \: d_{i,c} \: - \: d_{i,f} \: |^2 \:
\geq \: 0 \: .
\label{ez217a}
\end{eqnarray}
From Eq.(\ref{ez217a}), one can see that for all ${\vec k}$, the value of 
$P_{\vec k}$ is non-negative, i.e. ${\cal J} \: > \: 0$ \cite{E3}. So we have
\begin{eqnarray}
\langle \: \hat R_{loc} \: \rangle \: > \: E^U_0 \: , \quad 
\langle \: \hat R_{mov} \: \rangle \: < \: 0 \: .
\label{ez218}
\end{eqnarray}
Taking into consideration Eq.(\ref{ez218}), it can be seen that the 
ground-state energy cannot be expressed as a sum over expectation values of 
on-site contribution terms of the Hamiltonian. We have $\langle \: \hat 
R_{loc} \: \rangle \: > \: E^U_g $, and as a consequence, the system is not 
localized. From the other side, since $\langle \: \hat R_{mov} \: \rangle \:
< \: 0 \: $, the sum of the expectation values of Hamiltonian terms that 
preserve the movement of the particles within the system, is nonzero and 
negative. As a consequence, the ground-state energy is exactly $E^U_g$, 
because in this way the system is maintaining its itinerant character that 
allows to reach the state with the minimum possible energy. In this 
conditions, the ground-state is a 2D normal state (i.e. non-symmetry 
broken) NFL, which is paramagnetic and non-insulating. 
The presented ground-state expectation values are correct only for $U \: >
\: 0$, since for the noninteracting $U \: = \: 0$ case, the $| \: \Psi^U_g \:
\rangle$ contained in Eq.(\ref{ez17}) represents only a negligible fraction 
from the linear combination of wave functions that build up the ground-state
$| \: \Psi^0_g \: \rangle$ defined at the end of Sec.II. via the arbitrary 
operator $\hat Q$ instead of $\hat F^{(3)}$. As a consequence, we can clearly
state, that the interacting ground-state cannot be obtained perturbatively 
from the ground-state of the $U \: = \: 0$ case. This confirms the general 
belief, that a NFL emergence in normal phase and two dimensions
has to be a completely non-perturbative effect, similar to 1D case \cite{fuk}.

Doping the system above $3 / 4$ filling, the ground-state wave function
becomes $| \: \Psi_{g,d}^U \: \rangle$ from Eq.(\ref{ez21}). With the operator
$\hat F^{(4)}$ in $| \: \Psi^U_{g,d} \: \rangle$, given by the product 
$\prod_{\beta = 1}^3 \: \hat F^{(\beta)}$ present in the ground-state wave
function, we further have $[ \: \hat G \: + \: U \: \hat P' \: ] \: | \:
\Psi^U_{g,d} \: \rangle \: = \: 0$. The band remaining flat and being not 
possible to arrange the particles in such a way, to have the same number of 
electrons on every site of the lattice, the system is not localized, and 
remains a non-Fermi liquid as well. 

\subsection{Excited states}

Starting from Eq.(\ref{ez203}), it can be seen that excited states are 
obtained by removing $\hat B^{\dagger}_{{\vec k}, 2,\sigma}$
operators from the first product of Eq.(\ref{ez203}), changing their band 
index to $1$ (i.e. removing particles from the completely filled lower band
to the partially filled upper band), and leaving intact the $F^{(3)}$ 
component of the wave function. For example, removing one $\hat B^{\dagger}_{
{\vec k}, 2, - \sigma}$, we obtain
\begin{eqnarray}
| \: \Psi^U_{\vec k_1} \: \rangle \: = \: \left[ \: \prod_{\vec k} \:
\hat B^{\dagger}_{{\vec k}, 2, \sigma} \: \right] \: \left[ \: 
\prod_{{\vec k} \ne {\vec k_1}} \: \hat B^{\dagger}_{{\vec k}, 2, -\sigma}
\: \right] \: ( \: \hat B^{\dagger}_{{\vec k_1}, 1, -\sigma} \: ) \:
\hat F^{(3)} \: | \: 0 \: \rangle \: .
\label{ex1}
\end{eqnarray}
Since in Eq.(\ref{ex1}) the $\hat F^{(3)}$ operator remains intact, it further
preserves at least one $f$-electron on every site of the lattice, so
$\hat P' \: | \: \Psi^U_{\vec k_1} \: \rangle \: = \: 0$ also in this case. 
Using Eq.(\ref{ez208c}), the remaining part of the Hamiltonian besides 
$\hat P'$ is essentially $\sum_{{\vec k}, \sigma} \: ( \: E_{{\vec k}, 1} \:
\hat B^{\dagger}_{{\vec k}, 1, \sigma} \: \hat B_{{\vec k}, 1, \sigma} \:
+ \: E_{{\vec k}, 2} \: \hat B^{\dagger}_{{\vec k}, 2, \sigma} \:
\hat B_{{\vec k}, 2, \sigma} \: )$, for which, the component of
$| \: \Psi^U_{\vec k_1} \: \rangle$ from Eq.(\ref{ex1}) 
which is orthogonal to the ground-state
represents an eigenstate. Other eigenstates of 
this type can be obtained removing more than one $\hat B^{\dagger}_{{\vec k},
2, \sigma}$ operator from the lower band and changing their band index to 
$1$. A similar procedure can be applied also in the presence of doping. These 
excited states (for example, for one removed $\hat B^{\dagger}_{{\vec k}, 2,
\sigma}$), have energy $E^{(1)}_{{\vec k_1}}$ greater than the ground-state 
energy with a value of order $\Delta \: = \: Min [ \: \Delta_{\vec k} \: ]$. 
Since in the analyzed case $\Delta_{\vec k} \: > \: 0$, the obtained energy 
spectrum for Eq.(\ref{ex1}) is gaped.

The excited states of the type presented in Eq.(\ref{ex1}) were obtained by
modifications in the first two operatorial components $\hat F^{(1)}$ and
$\hat F^{(2)}$ of the ground-state wave function from Eq.(\ref{ez17}). In 
principle, is not possible to exclude excited states obtained by modifications
made at the level of $\hat F^{(3)}$ part of the wave function, by introducing 
in it a double occupancy - empty site pair in direct space (creating in fact 
a supplementary double occupied site at the level of $| \: \Psi^U_g \:
\rangle$ from Eq.(\ref{ez17}).). In this case, the decompositions used for 
$\hat H$ presented in Eqs.(\ref{ez6},\ref{ez7}) are not usable, and even 
the band structure described by $E_{{\vec k}, i}$, $i \: = \: 1, 2$ in 
Eq.(\ref{ez40}) is questionable at $U \: \ne \: 0$. However, these excited 
states have energy greater than the ground-state energy with a value of 
order ${\cal O}(U)$, since a supplementary double occupied state has been 
introduced in the direct space on a given lattice site. Taking 
\begin{eqnarray}
U \: >> \: \Delta \: ,
\label{ex2}
\end{eqnarray}
the low lying excitation spectrum of the system will be clearly dominated by
$E^{(1)}_{{\vec k_1}}$ given by the states presented in Eq.(\ref{ex1}). This
means, that the low lying excitation spectrum will be gaped, so will be
clearly visible in the physical properties of the system at $T \: > \: 0$. 
The gap becomes $\Delta_{\vec k}$ (i.e. is ${\vec k}$ dependent), its minimum
value being $\Delta$. The gap symmetry is a possible symmetry allowed by the 
described 2D lattice and depends on the starting parameters of the system via
Eqs.(\ref{ez26},\ref{ez53}).  

\section{Summary and Conclusions} 

In conditions in which even the exact solution for the 1D periodic Anderson
model (PAM) is not known at finite $U$, we present in this paper 
rigorous and exact solutions for 2D - PAM in the interacting case.
The described solutions are present on two surfaces of the $T \: = \: 0$ 
parameter space of the model. Both solutions describe the interacting 
$U \: > \: 0$ model, and the deduced ground-state wave functions cannot be 
obtained perturbatively from the non-interacting case.

The first solution described in Sec.V. emerges for pure real hybridization 
coupling constants at $3 / 4$ filling and represents a paramagnetic Mott 
insulator, the ground-state being completely localized. 

The second solution presented in Sec. VI. represents a new non-Fermi liquid 
in 2D normal (non-symmetry broken) phase and emerges in case of pure 
imaginary hybridization coupling constants (the Hamiltonian remaining 
hermitian) at $N / N_{\lambda} \: \geq \: 3 / 4$ filling. This 
phase presents an $n_{\vec k}$ momentum distribution function which is 
continuous together with its derivatives of any order, has a well defined 
Fermi energy, but the Fermi momentum is not definable and the system has no 
Fermi surface in the ${\vec k}$-space. At high on-site repulsion $U$, the 
described phase has a gap in the density of low lying excitations, and 
in the parameter space it emerges in the vicinity of a Mott insulating phase.
The ground-state presents a large spin degeneracy and is paramagnetic. The 
behavior physically is given by a partially filled upper completely flat band
situated above a normal band with dispersion. Low lying excitations do not 
give quasi-particles above the Fermi level, and have the effect to increase 
the number of particles at the Fermi energy. In this process particles 
are removed from the lower band and introduced into the upper flat band.  

We have to underline that in the case of non-distorted unit cell, from the
presented ground-state solutions only the completely localized state emerges.
Introducing distortions in the unit cell (maintaining however its translation
invariance) we allow in fact the emergence of the described non-Fermi liquid 
state in the normal phase. In this way, the presented non-Fermi liquid
phase is strongly related to the presence of lattice distortions.

\acknowledgements

The authors kindly acknowledge elevating discussions on the subject and
critical reading of the manuscript for P. de Chatel, M. Gulacsi,  
A.S. Alexandrov. For Zs. G. research supported by OTKA-T022874 and FKFP-0471,
and for P.G. by the Research Fellowship Award of the Institute of Nuclear
Research Debrecen.

\section*{Figures Captions}

Fig.1. The hopping amplitudes for c-electrons (A), and hybridization coupling
constants for $\hat c^{\dagger} \hat f$ type hybridization terms (B) 
connecting the nearest and next-nearest neighbors lattice sites. The hopping
amplitudes for f-electrons, and hybridization couplings for $\hat f^{\dagger}
\hat c $ type hybridization terms are similar.

Fig.2. The effect of the Hubbard interaction $U$ in the flattening of the
$E_{{\vec k}, 1, \sigma}$ band. The $U$ and $E_{{\vec k},1}$ values are
in $|t_{f, x+y}|$ units, and $k_x, \: k_y$ cover the first Brillouin zone.
For the values of the other Hamiltonian parameters see text.

\newpage


\appendix

\section{}

This Appendix contains the mathematical details related to the calculation of
ground-state expectation values in case of arbitrary, and site dependent
coefficients $\alpha_{{\bf i},\sigma}$ introduced by the operator 
$\hat F^{(3)}$ in Eq.(\ref{ez16}). The calculations are presented for the 
non-doped case.

First of all, we compute the one-particle ground state expectation values of
the form $ \hat \Theta_{b,b'} \: = \: \sum_\sigma \: \hat b^\dagger_{\vec k,
\sigma} \: \hat b'_{\vec k, \sigma} , $ where $b \: , \: b' \: = \: c \: , \:
f$, and we show that these are independent from the coefficients $\alpha_{{\bf
i},\sigma}$.

For shake of simplicity of the presentation we use the notation $\bfsigma \: =
 \: ( \: \sigma_{\bf i} \: )_{{\bf i}=1}^{N_\Lambda}$, $\sigma_{\bf i} \: \in
 \: \{ \: \uparrow , \: \downarrow \: \}$, and denote by ${\hat
F}^{(3)}_{\bfscrsigma}$ the ${\hat F}^{(3)}$ operator containing a concrete set
of values $\alpha_{{\bf i},\uparrow} \: = \: \delta_{\sigma_{\bf i},\uparrow}$
and $\alpha_{{\bf i},\downarrow} \: = \: \delta_{\sigma_{\bf i},\downarrow}$
(i.e. zero or one depending on $\bfsigma$). We prove that the set of vectors
defined by all possible $\bfsigma$ values (see also Eq.(\ref{ez203}))
\begin{eqnarray}
| \: \Psi^U_{g,\bfscrsigma} \: \rangle \: = \: {\hat F}^{(1)} \: {\hat F}^{(2)}
\: {\hat F}^{(3)}_{\bfscrsigma} \: | \: 0 \: \rangle \: = \: \left[ \:
\prod_{\vec k} \: \left( \: \hat B^{\dagger}_{{\vec k},2,\uparrow} \: \hat
B^{\dagger}_{{\vec k},2,\downarrow} \: \right) \: \right] \: \hat
F^{(3)}_{\bfscrsigma} \: | \: 0 \: \rangle 
\label{a1}
\end{eqnarray}
give the same expectation values for $\hat \Theta_{b,b'}$ independent on
$\bfsigma$. We mention that the wave vectors presented in Eq.(\ref{a1}) 
span the subspace of the ground state, they are non-orthogonal, but as 
seen below are linearly independent.

We express also the operator $F^{(3)}_{\bfscrsigma}$ in terms of $\hat
B^\dagger_{\vec k,j,\sigma}$. For this reason first the Fourier transforms of
the original $\hat b_{{\bf i},\sigma}$ operators, then the Eq.(\ref{ez208b})
must be used.  Due to the first term in Eq.(\ref{a1}) every $\hat
B^{\dagger}_{{\vec k},2,\sigma}$ from $\hat F^{(3)}_{\bfscrsigma}$ cancels out
and we find
\begin{eqnarray}
| \: \Psi^U_{g,\bfscrsigma} \: \rangle \: & = & \: \left[ \: \prod_{\vec k} \:
\left( \hat B^{\dagger}_{{\vec k},2,\uparrow} \: \hat B^{\dagger}_{{\vec
k},2,\downarrow} \: \right) \: \right] \: \prod_{\bf i} \: \left( \:
\sum_{{\vec k}_{\bf i}} \: e^{- i \vec k_{\bf i} \: \vec r_{\bf i}} \: \frac{
- \: X_c( \: {\vec k}_{\bf i} \: ) \: \hat B^{\dagger}_{{\vec k}_{\bf i},1,
\sigma}}{\sqrt{ \: \Delta_{\vec k_{\bf i}}}}
 \: \right) \: | \: 0 \: \rangle 
\nonumber \\
& = & \: \left[ \: \prod_{\vec k} \: \left( \: \hat B^{\dagger}_{{\vec
k},2,\uparrow} \: \hat B^{\dagger}_{{\vec k},2,\downarrow}
 \: \right) \: \right] \: \left[ \: \sum_{\bf k} \: \delta_{S^z( {\bf k}  ) ,
S^z(\bfscrsigma)} \: Z( \: {\bf k} \: ) \: \bar Y( \: {\bf k},\bfsigma \: ) \: 
\hat F_{{\bf k},1} \: \right] \: | \: 0 \: \rangle \: .
\label{a6}
\end{eqnarray}
Here ${\bf k}$ denotes a set of ${\vec k}$ values whose first $N_{\uparrow}$
elements (i.e. ${\vec k}_i$ for spin-up states) and also the last
$N_{\downarrow} \: = \: N_{\Lambda} \: - \: N_{\uparrow}$ elements
(i.e. ${\vec k}_i$ for spin-down states) are ordered, but there is no any
relation between spin-up and spin-down ${\vec k}$ values. As a consequence
${\bf k} \: = \: \{ \: ( \: \vec k_1 \: < \: \vec k_2 \: < \: \dots \: < \:
\vec k_{N_\uparrow} \: ) \: , \: ( \: \vec k_{N_\uparrow + 1} \: < \: \vec
k_{N_\uparrow + 2} \: < \: \dots \: < \: \vec k_{N_\Lambda} \: ) \: \}$, and
$\sum_{\bf k} $ means a sum over all different $\{ \: {\vec k} \: \}$
configurations allowed by $\bfsigma$ (i.e. $S^z( \: {\bf k} \: ) \: = \: ( \:
N_\uparrow \: - \: N_\downarrow \: ) \: / \: 2$ and $S^z( \: \bfsigma \: ) \:
= \: \sum_{\bf i} \: \frac{1}{2} \: ( \: \delta_{\sigma_{\bf i},\uparrow} \: -
\: \delta_{\sigma_{\bf i},\downarrow} \: )$ are equal). Denoting $N_{\uparrow}
\: = \: N_{\Lambda} \: / \: 2 \: + \: S^z( \: {\bf k} \: )$ we have
\begin{eqnarray} 
\hat F_{{\bf k},1} \: = \: \prod_{j \leq \frac{N_\Lambda}{2} + 
S^z( \: {\bf k} \: )} \: \hat B^{\dagger}_{{\vec k}_{j},1,\uparrow} \: 
\prod_{j > \frac{N_\Lambda}{2} + S^z( \: {\bf k} \: )} \: 
\hat B^{\dagger}_{{\vec k}_{j},1,\downarrow}, \: \quad \: Z( \: {\bf k} \: ) \:
= \: \prod_{j=1}^{N_\Lambda} \: \frac{- \: X_c( \: {\vec k}_{j} \: )}{
\sqrt{ \: \Delta_{\vec k_j} \: }} \: ,
\label{a8} 
\end{eqnarray} 
\begin{eqnarray} 
\bar Y( \: {\bf k} , \bfsigma \: ) \: = \: ( \: - \: 1 \: )^{ | Q | } \: 
\sum_{P \in G } \: ( \: - \: 1 \: )^{ | P | } \: \exp( \: i \: 
\sum_{j=1}^{N_\Lambda} \: \vec k_{P(j)} \: \vec r_{Q(j)} \: ) \: .
\label{a9} 
\end{eqnarray} 
In Eq.(\ref{a9}) $ | Q | $ is the parity of the permutation $Q$ producing the
spin-up spin-down separation (i.e. $Q( \: \bfsigma \: ) \: = \: ( \:
\sigma_{Q( \: {\bf i} \: )} \: )_{{\bf i}=1}^{N_\Lambda} \: = \: \{ \:
\uparrow , \: \uparrow , \: \dots , \: \uparrow , \: \downarrow , \:
\downarrow, \: \dots , \: \downarrow \: \} \: $), and $ | P | $ is the parity
of the permutation $P \: \in \: G \: = \: S_{\{1, \dots , N_{\uparrow} \} } \:
\times S_{ \{ N_{\uparrow} + 1 , \dots , N_\Lambda \} }$ which rearranges the
${\vec k}$ subsets ${\uparrow}$ and ${\downarrow}$.

On the other hand, the Fourier transform of $\hat F^{(3)}_{\bfscrsigma} \: =
\: \prod_{\bf i} \: \hat f^{\dagger}_{{\bf i} , \sigma_{\bf i}}$ gives
\begin{eqnarray}
{\hat F}^{(3)}_{\bfscrsigma} \: | \: 0 \: \rangle \: = \: 
\sum_{{\bf k}} \: 
\delta_{S^z( {\bf k}) , S^z( \bfsigma )} \: Y( \: {\bf k} , \bfsigma \: ) \:  
\prod_{j \leq \frac{N_\Lambda}{2} + S^z( {\bf k} )} \: 
 \hat f^{\dagger}_{{\vec k}_{j},\uparrow} \: 
\prod_{j > \frac{N_\Lambda}{2} + S^z( {\bf k} )} \:  
 \hat f^{\dagger}_{{\vec k}_{j},\downarrow} \: 
| \: 0 \: \rangle \: ,
\label{a10}
\end{eqnarray}
which excepting $Z( \: {\bf k} \: )$ is exactly the second parentheses from
Eq.(\ref{a6}). Based on this observation it can be proved that the
non-orthogonal set of ground-state wave-vectors from Eq.(\ref{a1}) are
linearly independent. In order to do this, we prove that there exists a dual
basis (i.e., $| \: \Psi^U_{g,\bfscrsigma} \: \rangle$ are linearly
independent). The dual basis has the form
\begin{eqnarray}
| \: \Psi^{U,\bfscrsigma}_{g} \: \rangle \: = \:
\left[ \: \prod_{\vec k} \: \left( \: 
\hat B^{\dagger}_{{\vec k},2,\uparrow} \: \hat B^{\dagger}_{{\vec k},2,
\downarrow} \: \right) \: \right] \: \sum_{\bf k} \: \delta_{S^z( {\bf k} ) ,
S^z( \bfsigma) } \: 
\frac{1}{Z^*( \: {\bf k} \: )} \: Y( \: {\bf k} , \bfsigma \: ) \: 
\hat F_{{\bf k},1} \: | \: 0 \: \rangle \: ,
\label{a11}
\end{eqnarray}
and the proof via Eq.(\ref{a10}) is simple 
\begin{eqnarray}
&& \langle \: \Psi^{U,\bfscrsigma^{\prime}}_{g} \: | \: 
\Psi^U_{{g,\bfscrsigma}} \: \rangle \: =
\nonumber \\
&& = \: \sum_{{\bf k}^{\prime}} \: \sum_{{\bf k}} \: 
\delta_{S^z( {\bf k}^{\prime}) , S^z(\bfscrsigma^{\prime} )} \: 
\frac{1}{Z( \: {\bf k}^{\prime} \: )} 
Y^*( \: {\bf k}^{\prime} , \bfsigma^{\prime} \: ) \: 
\delta_{S^z( {\bf k} ) , S^z( \bfscrsigma )} \: 
Z( \: {\bf k} \: ) \: Y( \: {\bf k} , \bfsigma \: ) \: 
\langle \: 0 \: | \: \hat F^{\dagger}_{{\bf k}^{\prime},1} \: 
\hat F_{{\bf k},1} \: | \: 0 \: \rangle
\nonumber \\
&& = \: \sum_{{\bf k}} \: \delta_{S^z( {\bf k} ) , S^z( \bfscrsigma^{\prime} )}
\: \delta_{S^z( {\bf k} ) , S^z( \bfscrsigma )} \: 
Y^*( \: {\bf k} , \bfsigma^{\prime} \: ) \: Y( \: {\bf k} , \bfsigma \: ) \: 
= \: \sum_{{\bf k}} \: 
\delta_{S^z( \bfscrsigma ) , S^z( \bfscrsigma^{\prime} )} \: 
Y^*( \: {\bf k} , \bfsigma^{\prime} \: ) \: Y( \: {\bf k} , \bfsigma \: )
\nonumber \\
&& = \: \sum_{{\bf k}^{\prime}} \: \sum_{{\bf k}} \: 
\delta_{S^z( {\bf k}^{\prime} ) , S^z( \bfscrsigma^{\prime} )} \: 
Y^*( \: {\bf k}^{\prime} , \bfsigma^{\prime} \: ) \: 
\delta_{S^z( {\bf k} ) , S^z( \bfscrsigma )} \: 
Y( \: {\bf k} , \bfsigma \: ) \: \times 
\nonumber \\ &&
\langle \: 0 \: | \: 
\Bigl( \: \prod_{j \leq \frac{N_\Lambda}{2} + S^z( {\bf k}^\prime )} \: 
\hat f^{\dagger}_{{\vec k}^{\prime}_{j},\uparrow} \: 
\prod_{j > \frac{N_\Lambda}{2} + S^z( {\bf k}^\prime )} \: 
\hat f^{\dagger}_{{\vec k}^{\prime}_{j},\downarrow} \: \Bigr)^\dagger 
\Bigl( \: \prod_{j \leq \frac{N_\Lambda}{2} + S^z( {\bf k} )} \: 
\hat f^{\dagger}_{{\vec k}_{j},\uparrow} \: 
\prod_{j > \frac{N_\Lambda}{2} + S^z( {\bf k} )} \: 
\hat f^{\dagger}_{{\vec k}_{j},\downarrow} \: \Bigr) \: | \: 0 \: \rangle 
\nonumber \\
&& = \: \langle \: 0 \: | \: {\hat F}^{(3){\dagger}}_{\bfscrsigma^{\prime}} \: 
{\hat F}^{(3)}_{\bfscrsigma} \: | \: 0 \: \rangle \: 
= \: \delta_{\bfscrsigma,\bfscrsigma^{\prime}}
\label{a12}
\end{eqnarray}
Due to the fact that all $\hat F_{{\bf k},1} \: | \: 0 \: \rangle$ are
eigenvectors of the operator $\hat B^{\dagger}_{{\vec k},1,\sigma} \: \hat
B_{{\vec k},1,\sigma}$ with the same eigenvalue than the
$\prod_{j \leq \frac{N_\Lambda}{2} + S^z( {\bf k} )} \: 
\hat f^{\dagger}_{{\vec k}_{j},\uparrow} \: 
\prod_{j > \frac{N_\Lambda}{2} + S^z( {\bf k} )} \: 
\hat f^{\dagger}_{{\vec k}_{j},\downarrow} \: 
| \: 0 \: \rangle$ eigenvectors of the operator  
$\hat f^{\dagger}_{{\vec k},\sigma} \: \hat f_{{\vec k},\sigma}$,
similarly to Eq.(\ref{a12}) we find
\begin{eqnarray}
&& \langle \: \Psi^{U,\bfscrsigma^{\prime}}_{g} \: | \: 
\sum_{\sigma} \: \hat B^{\dagger}_{{\vec k},1,\sigma} \: \hat B_{{\vec
k},1,\sigma} \: 
| \: \Psi^U_{{g,\bfscrsigma}} \: \rangle \: 
= \: \langle \: 0 \: | \: {\hat F}^{(3){\dagger}}_{\bfscrsigma^\prime} \: 
\sum_{\sigma} \: \hat f^{\dagger}_{{\vec k},\sigma} \: \hat f_{{\vec
k},\sigma} \: 
{\hat F}^{(3)}_{\bfscrsigma} \: | \: 0 \: \rangle \: 
\nonumber \\
&& = \: \frac{1}{N_\Lambda} \: \sum_{{\bf j, j}^\prime} \: 
e^{ - i \vec k (\vec r_{\bf j} - \vec r_{{\bf j}^\prime} ) } \:
 \sum_{\sigma} \: 
\langle \: 0 \: | \: \hat F^{(3) \: \dagger}_{\bfscrsigma^\prime} \: 
\hat f^{\dagger}_{{\bf j},\sigma} \: \hat f_{{\bf j}^\prime,\sigma} \: 
\hat F^{(3)}_{\bfscrsigma} \: | \: 0 \: \rangle 
\nonumber \\
&& = \: \frac{1}{N_\Lambda} \: \sum_{{\bf j, j}^\prime} \: 
e^{ - i \vec k (\vec r_{\bf j} - \vec r_{{\bf j}^\prime} ) } \: 
\sum_{\sigma} \: \delta_{\sigma , \bfscrsigma_{\bf j}} \: 
\delta_{{\bf j}, {\bf j}^{\prime}} \: \delta_{\bfscrsigma, \bfscrsigma^\prime}
\: = \: \delta_{\bfscrsigma,\bfscrsigma^{\prime}} \: .
\label{a13}
\end{eqnarray}
The dual-basis to basis expectation value from Eq.(\ref{a13}) gives however 
\begin{eqnarray}
&& \langle \: \Psi^U_{g,\bfscrsigma} \: | \: 
\sum_{\sigma} \: \hat B^{\dagger}_{{\vec k},1,\sigma} \: \hat B_{{\vec
k},1,\sigma} \: 
| \: \Psi^U_{{g,\bfscrsigma}} \: \rangle \: = 
\nonumber\\
&& \sum_{\bfscrsigma^{\prime}} \: 
\langle \: \Psi^U_{g,\bfscrsigma} \: | \: \Psi^U_{g,\bfscrsigma^{\prime}} \: 
\rangle \: \langle \: \Psi^{U,\bfscrsigma^{\prime}}_{g} \: | \: 
\sum_{\sigma} \: \hat B^{\dagger}_{{\vec k},1,\sigma} \: 
\hat B_{{\vec k},1,\sigma} \: | \: \Psi^U_{{g,\bfscrsigma}} \: \rangle \: 
= \: \langle \: \Psi^U_{g,\bfscrsigma} \: | \: \Psi^U_{g,\bfscrsigma} \: 
\rangle \: ,
\label{a14}
\end{eqnarray}
from where we find that independent from $\bfsigma$ we have
\begin{eqnarray}
\frac{ \langle \: \Psi^U_{g,\bfscrsigma} \: | \: 
\sum_{\sigma} \: \hat B^{\dagger}_{{\vec k},1,\sigma} \: \hat B_{{\vec
k},1,\sigma} \: 
| \: \Psi^U_{{g,\bfscrsigma}} \: \rangle }{ \langle \: 
\Psi^U_{g,\bfscrsigma} \: | \: \Psi^U_{g,\bfscrsigma} \: \rangle } \: 
= \: 1 \: .
\label{a15}
\end{eqnarray}
After this step, using the form of the ground state wave function from
Eq.(\ref{a1}) we simply obtain for all $\sigma$ and $\bfsigma$
\begin{eqnarray}
\hat B^{\dagger}_{{\vec k},2,\sigma} \: \hat B_{{\vec k},2,\sigma} \: | \: 
\Psi^U_{g,\bfscrsigma} \: \rangle \: = \: | \: \Psi^U_{g,\bfscrsigma} \: 
\rangle \: , \quad 
\hat B^{\dagger}_{{\vec k},2,\sigma} \: \hat B_{{\vec k},1,\sigma} \: 
| \: \Psi^U_{g,\bfscrsigma} \: \rangle \: = \: 0 \: .
\label{a16}
\end{eqnarray}
From Eq.(\ref{a16}), $\langle \: \Psi^U_{g,\bfscrsigma} \: | \: \hat
B^{\dagger}_{{\vec k}, 2,\sigma} \: \hat B_{{\vec k},2,\sigma} \: | \:
\Psi^U_{g,\bfscrsigma} \: \rangle \: / \: \langle \: \Psi^U_{g,\bfscrsigma} 
\: | \: \Psi^U_{g,\bfscrsigma} \: \rangle \: = \: 1 \: ,$ and $\langle \:
\Psi^U_{g,\bfscrsigma} \: | \: \hat B^{\dagger}_{{\vec k},2,\sigma} \: \hat
B_{{\vec k},1,\sigma} \: | \: \Psi^U_{g,\bfscrsigma} \: \rangle \: = \: 0$
relations arise independent on $\bfsigma$ spin arrangement of the
ground-state.  Eq.(\ref{a16}) shows that in order to calculate from $\hat
\Theta_{b,b'}$ the desired expectation values $\langle \:
\Psi^U_{g,\bfscrsigma} \: | \: \sum_{\sigma} \: {\hat b}^{\dagger}_{{\vec k},
\sigma} \: {\hat b}^{\prime}_{{\vec k}, \sigma} \: | \: \Psi^U_{g,\bfscrsigma}
\: \rangle \: / \: \langle \: \Psi^U_{g,\bfscrsigma} \: | \:
\Psi^U_{g,\bfscrsigma} \: \rangle$, first of all we may use Eq.(\ref{ez208b})
to transform the initial operators into ${\hat B}^{\dagger}_{{\vec k}, j,
\sigma}$ ($ j \: = \: 1 \: , \: 2 $). After this step the non-diagonal $\hat
B^{\dagger}_{{\vec k},1,\sigma} \: \hat B_{{\vec k},2,\sigma}$ terms cancel
out based on Eq.(\ref{a16}), and the remaining parts of the expectation value
can be deduced from Eq.(\ref{a15}) and first relation of Eq.(\ref{a16}).  For
example in the case of $\hat c^{\dagger}_{{\vec k},\sigma} \: \hat c_{{\vec
k},\sigma}$ we find
\begin{eqnarray}
\lefteqn{
\frac{\langle \: \Psi^U_{g,\bfscrsigma} \: | \: \sum_\sigma \: 
{\hat c}^{\dagger}_{{\vec k},\sigma} \: {\hat c}_{{\vec k}, \sigma} \: | \: 
\Psi^U_{g,\bfscrsigma} \: \rangle}{ \langle \: \Psi^U_{g,\bfscrsigma} \: | \: 
\Psi^U_{g,\bfscrsigma} \: \rangle} \: =}
\nonumber \\
& = & \: \frac{1}{\Delta_{\vec k} \: \langle \: \Psi^U_{g,\bfscrsigma} \: 
| \: \Psi^U_{g,\bfscrsigma} \: \rangle } \: 
\langle \: \Psi^U_{g,\bfscrsigma} \: | \: \sum_\sigma \: \left( \: 
X_c( \: {\vec k} \: ) \: X_c^*( \: {\vec k} \: ) \: 
\hat B^{\dagger}_{{\vec k},2,\sigma} \: \hat B_{{\vec k},2,\sigma} \: +
\right. \nonumber \\ && \left.
+ \: X_f( \: {\vec k} \: ) \: X_f^*( \: {\vec k} \: ) \: 
\hat B^{\dagger}_{{\vec k},1,\sigma} \: \hat B_{{\vec k},1,\sigma} \: 
+ \: X_c^*( \: {\vec k} \: ) \: X_f^*( \: {\vec k} \: ) \:  
\hat B^{\dagger}_{{\vec k},2,\sigma} \: \hat B_{{\vec k},1,\sigma} \: +
\right. \nonumber \\ && \left.
+ \: X_c( \: {\vec k} \: ) \: X_f( \: {\vec k} \: ) \: 
\hat B^{\dagger}_{{\vec k},1,\sigma} \: \hat B_{{\vec k},2,\sigma} \: 
\right) \: | \: \Psi^U_{g,\bfscrsigma} \: \rangle 
\nonumber \\
& = & \: \frac{2 \: | \: X_c( \: {\vec k} \: ) \: |^2 \: + \: | \: X_f( \:
{\vec k} \: ) \: |^2 }
{ | \: X_c( \: {\vec k} \: ) \: |^2 \: + \: | \: X_f( \: {\vec k} \: ) \: |^2
} \: = \: 
\frac{2 \: | \: X_c( \: {\vec k} \: ) \: |^2 \: + \: | \: X_f( \: {\vec k} \:
) \: |^2 }{\Delta_{\vec k}} \: 
= \: 1 \: + \: \frac{ \: K \: - \: \epsilon^c_{\vec k} }{ \Delta_{\vec k} }
\: ,
\label{a17}
\end{eqnarray}
where we used Eq.(\ref{ez52}) and Eq.(\ref{ez53}). Similarly
\begin{eqnarray}
\frac{ \langle \: \Psi^U_{g,\bfscrsigma} \: | \: \sum_\sigma \: {\hat
f}^{\dagger}_{{\vec k}, \sigma} \: {\hat f}_{{\vec k}, \sigma} \: | \:
\Psi^U_{g,\bfscrsigma} \: \rangle}{ \langle \: \Psi^U_{g,\bfscrsigma} \: | \:
\Psi^U_{g,\bfscrsigma} \: \rangle} \: & = & \: 1 \: + \: \frac{ K \: - \:
\epsilon^f_{\vec k} }{ \Delta_{\vec k} } 
\nonumber \\
 \quad \frac{\langle \:
\Psi^U_{g,\bfscrsigma} \: | \: \sum_\sigma \: {\hat f}^{\dagger}_{{\vec k},
\sigma} \: {\hat c}_{{\vec k}, \sigma} \: | \: \Psi^U_{g,\bfscrsigma} \:
\rangle}{ \langle \: \Psi^U_{g,\bfscrsigma} \: | \: \Psi^U_{g,\bfscrsigma} \:
\rangle} \: & = & \: \frac{ - \: V_{\vec k} }{ \Delta_{\vec k} }
\label{a18}
\end{eqnarray}
With Eqs. (\ref{a17},\ref{a18}) the independence of the analyzed expectation 
values on $\alpha_{{\bf i},\sigma}$ has been demonstrated.

We concentrate now on ground-state expectation values from operatorial terms
not summed over the spin index $\sigma$. In conditions in which 
$| \: \Psi^U_{g,\bfscrsigma} \: \rangle$ build up a non-orthogonalized basis,
the best way for this is to consider the $T\to 0$ limit of temperature 
dependent expectation values. We have
\begin{eqnarray}
\lim_{T \rightarrow 0} \: 
\frac{ {\rm Tr} \: e^{- \beta \hat H} \: \hat A }
{ {\rm Tr} \: e^{- \beta \hat H} } \: 
& = & \: \frac{ {\rm Tr}_{{\cal H}_0} \: \hat A }
{ {\rm Tr}_{{\cal H}_0} \: \hat 1 } \: 
= \: \frac{ \sum_{\bfscrsigma} \: \langle \: \Psi^{U,\bfscrsigma}_g \: | \: 
\hat A \: | \: \Psi^U_{g,\bfscrsigma} \: \rangle}{
\sum_{\bfscrsigma} \: \langle \: \Psi^{U,\bfscrsigma}_g \: | \: 
\Psi^U_{g,\bfscrsigma} \: \rangle} 
\nonumber \\
& = & \: \frac{1}{2^{N_\Lambda}} \: \sum_{\bfscrsigma} \: 
\langle \: \Psi^{U,\bfscrsigma}_g \: | \: \hat A \: | \: 
\Psi^U_{g,\bfscrsigma} \: \rangle \: .
\label{a19}
\end{eqnarray} 
Based on Eq.(\ref{a19}) the ground-state expectation value of 
$\hat B^{\dagger}_{{\vec k},1,\sigma} \: \hat B_{{\vec k},1,\sigma}$ becomes
\begin{eqnarray}
\langle \: \hat B^{\dagger}_{{\vec k},1,\sigma} \: 
\hat B_{{\vec k},1,\sigma} \: \rangle \: 
& = & \: \frac{1}{2^{N_\Lambda}} \: \sum_{\bfscrsigma} \: 
\langle \: \Psi^{U,\bfscrsigma}_g \: | \: \hat B^{\dagger}_{{\vec k},1,\sigma} 
\: \hat B_{{\vec k},1,\sigma} \: | \: \Psi^U_{g,\bfscrsigma} \: \rangle
\nonumber \\
& = & \: \frac{1}{2^{N_\Lambda}} \: \sum_{\bfscrsigma} \: 
\langle \: 0 \: | \: {\hat F}^{(3){\dagger}}_{\bfscrsigma} \: 
 \hat f^{\dagger}_{{\vec k},\sigma} \: \hat f_{{\vec k},\sigma} \: 
{\hat F}^{(3)}_{\bfscrsigma} \: | \: 0 \: \rangle
\nonumber \\
& = & \: \frac{1}{2^{N_\Lambda}} \: \frac{1}{N_\Lambda} \: 
\sum_{{\bf j, j}^\prime} \: e^{ - i \vec k (\vec r_{\bf j} -
\vec r_{{\bf j}^\prime} ) }  \: \sum_{\bfscrsigma} \: 
\langle \: 0 \: | \: \hat F^{(3) \dagger}_{\bfscrsigma} \: 
\hat f^{\dagger}_{{\bf j},\sigma} \: \hat f_{{\bf j}^\prime,\sigma} \: 
 \hat F^{(3)}_{\bfscrsigma} \: | \: 0 \: \rangle
\nonumber \\
& = & \: \frac{1}{N_\Lambda} \: 
\sum_{{\bf j, j}^\prime} \: e^{ -i \vec k (\vec r_{\bf j} -
\vec r_{{\bf j}^\prime} ) } \: \delta_{{\bf j}, {\bf j}^{\prime}} \: 
 \frac{1}{2^{N_\Lambda}} \: \sum_{\bfscrsigma} \: 
\delta_{\sigma , \bfscrsigma_{\bf j}} \: 
= \: \frac{1}{2}
\label{a20}
\end{eqnarray} 
where the validity of the second equality is preserved by the same
argument as used for the first equality from Eq.(\ref{a13}), and the last 
equality holds because half of the all $\bfsigma$ has $\bfsigma_{\bf j} \: 
= \: \sigma$ value.

We mention, that the presented expectation values for 
$\hat B^{\dagger}_{{\vec k}, i, \sigma} \hat B_{{\vec k}, j, \sigma}$
operators are valid also for the complete ground-state wave function
$| \: \Psi^U_g \: \rangle \: = \: \sum_{\bfscrsigma} \: \gamma_{\bfscrsigma} 
\: | \: \Psi^{U}_{g,\bfscrsigma} \: \rangle$, where $\gamma_{\bfscrsigma}$ are
numerical coefficients. For the case of Eq.(\ref{a16}) this is trivial, and
in the case of Eq.(\ref{a19}) automatically all contributions 
$| \: \Psi^U_{g,\bfscrsigma} \: \rangle$ are taken into account. For the
the case of Eq.(\ref{a15}), using Eq.(\ref{a13}), we have
\begin{eqnarray}
&& \langle \: \Psi^U_{g} \: | \: 
\sum_{\sigma} \: \hat B^{\dagger}_{{\vec k},1,\sigma} \: \hat B_{{\vec
k},1,\sigma} \: 
| \: \Psi^U_{g} \: \rangle \: = 
\nonumber\\
&& \sum_{\bfscrsigma^{1},\bfscrsigma^{2},\bfscrsigma^{3}} \:
\gamma^{*}_{\bfscrsigma^{1}} \gamma_{\bfscrsigma^{2}} 
\langle \: \Psi^U_{g,\bfscrsigma^{1}} \: | \: \Psi^U_{g,\bfscrsigma^{3}} \: 
\rangle \: \langle \: \Psi^{U,\bfscrsigma^{3}}_{g} \: | \: 
\sum_{\sigma} \: \hat B^{\dagger}_{{\vec k},1,\sigma} \: 
\hat B_{{\vec k},1,\sigma} \: | \: \Psi^U_{{g,\bfscrsigma^{2}}} \: \rangle \: 
= 
\nonumber\\
&& \sum_{\bfscrsigma^{1},\bfscrsigma^{2},\bfscrsigma^{3}} \:
\gamma^{*}_{\bfscrsigma^{1}} \gamma_{\bfscrsigma^{2}} 
\langle \: \Psi^U_{g,\bfscrsigma^{1}} \: | \: \Psi^U_{g,\bfscrsigma^{3}} \: 
\rangle \:  \delta_{\bfscrsigma^{3},\bfscrsigma^{2}} \: =
\nonumber\\
&& \sum_{\bfscrsigma^{1},\bfscrsigma^{2}} \:
\gamma^{*}_{\bfscrsigma^{1}} \gamma_{\bfscrsigma^{2}} 
\langle \: \Psi^U_{g,\bfscrsigma^{1}} \: | \: \Psi^U_{g,\bfscrsigma^{2}} \: 
\rangle \: =
\: \langle \: \Psi^U_{g} \: | \: \Psi^U_{g} \: 
\rangle \: ,
\label{a14x}
\end{eqnarray}
from where $\langle \: \Psi^U_{g} \: | \: \sum_{\sigma} \: \hat
B^{\dagger}_{{\vec k}, 1 ,\sigma} \: \hat B_{{\vec k}, 1 ,\sigma} \: | \:
\Psi^U_{g} \: \rangle \: / \: \langle \: \Psi^U_{g} 
\: | \: \Psi^U_{g} \: \rangle \: = \: 1$ arises. 



\begin{references}
\bibitem{E1}
With mathematical rigorousity this means $\hat H | \Psi^U_g \rangle =
\hat H^{eff}_0 | \Psi^U_g \rangle$, where $\hat H^{eff}_0$ containing the
$U$ parameter as well, has the functional form of $\hat H_0$.
\bibitem{E3}
In the phase diagram we are situated now in conditions presented by 
Sec.VI.A.2., for which $P_{\vec k}$ becomes
$P_{\vec k} = |X^{*}_c(\vec k) + p X^{*}_f(\vec k)|^2 (|a_{1,f}|^2 + 
|a_{2,f}|^2) + |X^{*}_c(\vec k) - p X^{*}_f(\vec k)|^2 (|a_{3,f}|^2 +
|a_{4,f}|^2)$. The $X_c(\vec k)$ given in Eq.(\ref{ez152}) is not identically
zero, so clearly ${\cal{J}} > 0$ holds.
\end{references}
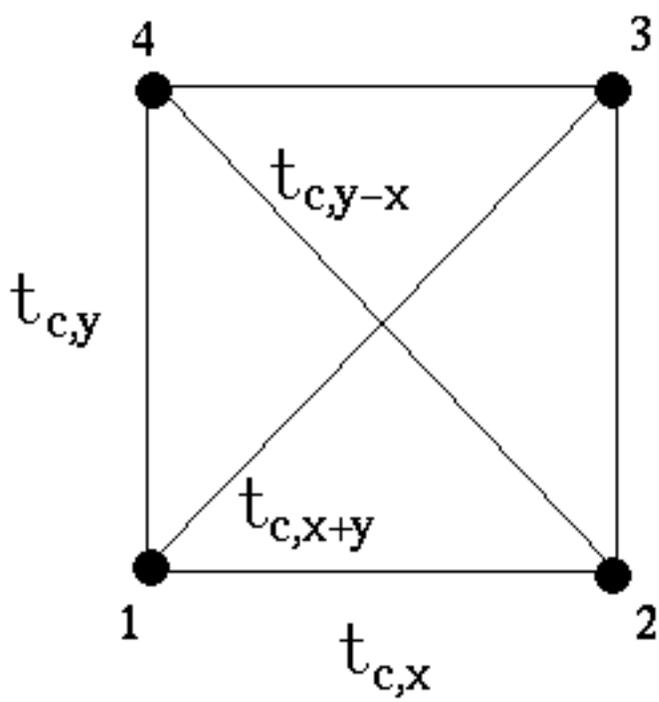
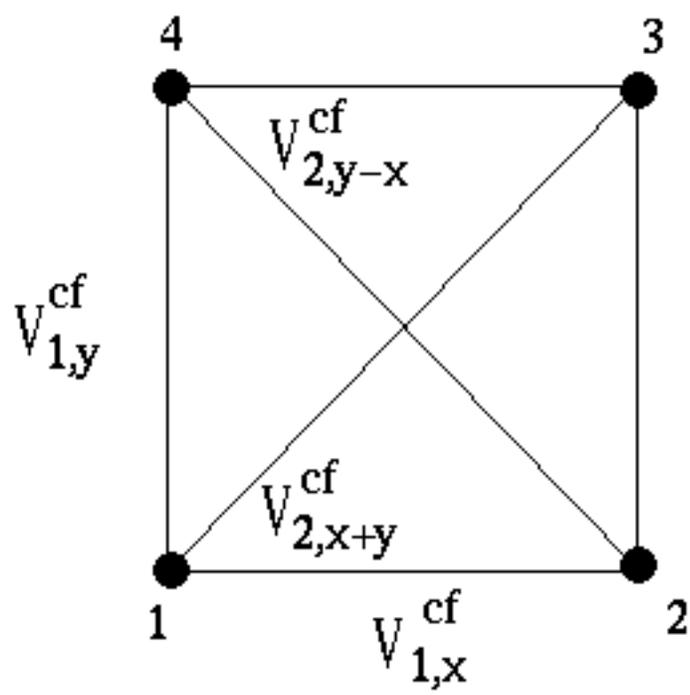
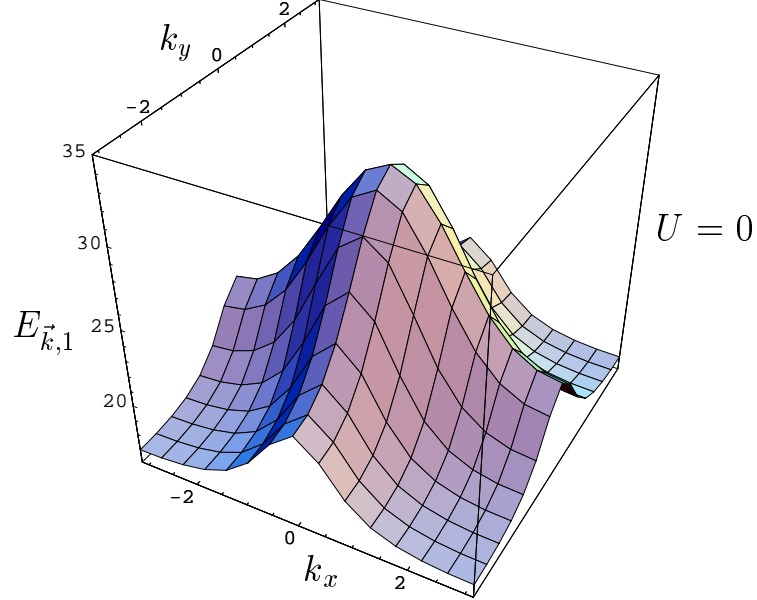
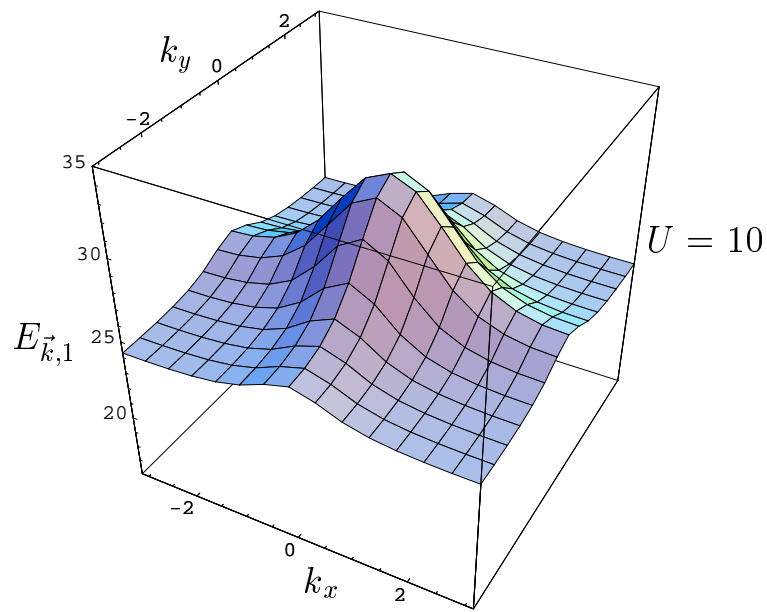
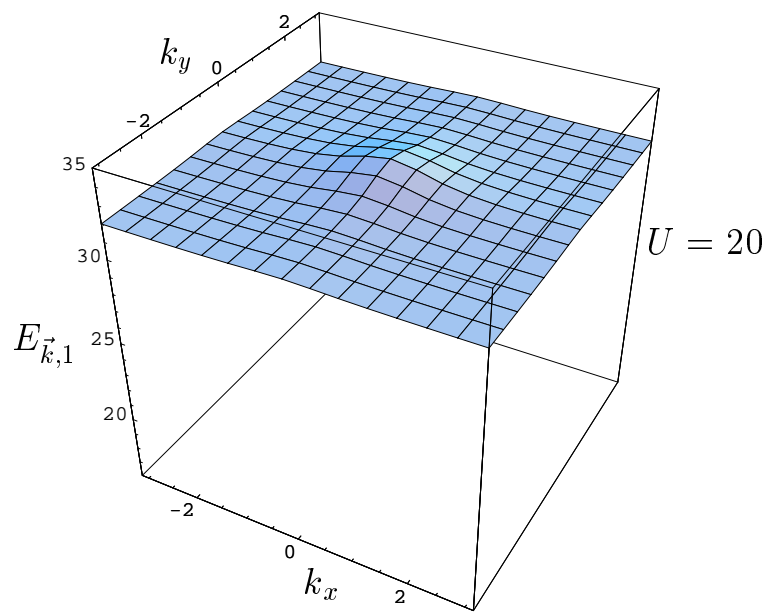
\end{document}